\documentclass[12pt,letterpaper]{JHEP3}

\usepackage{amssymb,amsmath,mathrsfs,slashed}

\newcommand{\Slash}[1]{{\ooalign{\hfil/\hfil\crcr$#1$}}}
\newcommand{\uD}{\mathrm{D}} \newcommand{\ve}{\varepsilon}
\newcommand{\vol}{V}
 \DeclareMathOperator*{\Tr}{Tr}

\title{Non-supersymmetric infrared perturbations to the warped
  deformed conifold}

\author{Paul McGuirk,$^{1}$ Gary Shiu,$^{1,2}$ and Yoske
  Sumitomo$^{1,2}$
  \\
  $^1$Department of Physics, University of Wisconsin, Madison, WI 53706, USA\\
  $^2$Institute for Advanced Study, Hong Kong University of Science
  and
  Technology, Hong Kong, People's Republic of China\\
  \email{mcguirk@physics.wisc.edu, shiu@physics.wisc.edu,
    sumitomo@wisc.edu}}

\date{\today}

\abstract{We analyze properties of non-supersymmetric
  isometry-preserving perturbations to the infrared region of the
  warped deformed conifold, i.e. the Klebanov-Strassler solution.  We
  discuss both perturbations that ``squash'' the geometry, so that the
  internal space is no longer conformally Calabi-Yau, and
  perturbations that do not squash the geometry. Among the
  perturbations that we discuss is the solution that describes the
  linearized near-tip backreaction of a smeared collection of
  $\overline{\uD 3}$-branes positioned in the deep infrared.  Such a
  configuration is a candidate gravity dual of a non-supersymmetric
  state in a large-rank cascading gauge theory.  Although
  $\overline{\uD 3}$-branes do not directly couple to the $3$-form
  flux, we argue that, due to the presence of the background imaginary
  self-dual flux, $\overline{\uD 3}$-branes in the Klebanov-Strassler
  geometry necessarily produce singular non-imaginary self-dual flux.
  Moreover, since conformally Calabi-Yau geometries cannot be
  supported by non-imaginary self-dual flux, the $\overline{\uD
    3}$-branes squash the geometry as our explicit solution shows.  We
  also briefly discuss supersymmetry-breaking perturbations at large
  radii and the effect of the non-supersymmetric perturbations on the
  gravitino mass.}

\preprint{MAD-TH-09-08}


\begin{document}

%%%%%
\section{Introduction}
%%%%%

Among the challenges in connecting string theory to our observable
universe is the difficulty in constructing controllable
supersymmetry-breaking backgrounds.  While supersymmetry (SUSY)
breaking is a prerequisite in any phenomenological study of
four-dimensional supersymmetric theories, the myriad of string theory
moduli makes this a formidable task.  Unless all moduli are stabilized
at a hierarchically higher scale than the scale of SUSY breaking, one
generically finds runaway directions that destabilize the vacuum,
taking us away from the controllable background which describes the
original supersymmetric state.

On top of this challenge, the observational evidence of an
accelerating universe adds yet another layer of complication: in
addition to the requirement that the SUSY-breaking background be
(meta)stable, viable vacua must also have positive energy density.
Motivated by this cosmological consideration, several mechanisms to
``uplift'' the vacuum energy of string vacua have since been
suggested, e.g., by adding ${\overline{\uD 3}}$-branes
\cite{Kachru:2003aw}, by introducing D-terms from gauge fluxes
\cite{Burgess:2003ic}, or by considering negatively curved internal
spaces
\cite{Silverstein:2007ac,Hertzberg:2007wc,Haque:2008jz,Danielsson:2009ff}
(see also~\cite{Caviezel:2008tf,Flauger:2008ad,deCarlos:2009fq}).
Though these mechanisms are often discussed in terms of 4D effective
field theories, it is of interest for a variety of reasons discussed
below to find backreacted supergravity solutions including such
uplifting sources as full 10D backgrounds.

In this paper, we report on some properties of non-supersymmetric
perturbations to the Klebanov-Strassler (KS) solution
\cite{Klebanov:2000hb}, a prototypical warped supersymmetric
background which is dual to a cascading $SU(N+M)\times SU(N)$ gauge
theory in the strong 't~Hooft limit, and is ubiquitous in flux
compactifications and in describing moduli stabilization.  The
backreaction of a collection of $\overline{\uD 3}$-branes placed at
the tip of the deformed conifold should be describable by such
perturbations. Such a configuration is known to be metastable against
brane/flux annihilation provided that the number of $\overline{\uD
  3}$-branes is sufficiently small in comparison to the background
flux~\cite{Kachru:2002gs}.  Though further instabilities generically
arise upon compactification when the closed string degrees of freedom
become dynamical and further stabilization mechanisms (e.g., fluxes,
non-perturbative effects, etc) are needed, this {\it local}
construction represents progress towards a genuine metastable
SUSY-breaking background.  Other than being an essential feature in
\cite{Kachru:2003aw} for vacuum uplifting to de Sitter space, the
warped $\overline{\uD 3}$ tension introduces an exponentially small
supersymmetry breaking scale which can be useful for describing hidden
sector dynamics (both in dimensionally reduced theories and in their
holographic descriptions).

Although we are interested especially in modes related to
$\overline{\uD3}$-branes, the analysis with more general modes brings
us interesting features for the classification of near-tip
perturbations.  We analyze perturbations that are either singular or
regular and those that either do or do not ``squash'' the geometry
(i.e. those that do or do not leave the internal geometry as
conformally the deformed conifold) in accordance with the equations of
motion.  We also identify which modes can break supersymmetry.  The
mode related to $\overline{\uD3}$s at the tip should have singular
behavior, at least in the warp factor, in order to capture the
localized tension.  We show below however that the only singular,
non-squashed, non-SUSY mode corresponds to a point source for the
dilaton, and thus cannot be identified as an $\overline{\uD 3}$-brane.
Furthermore, the squashed backreaction of an $\overline{\uD 3}$-brane
is supported by a $3$-form flux that is no longer imaginary self-dual
(ISD).  The fact that an $\overline{\uD 3}$-brane squashes the
geometry was observed in~\cite{DeWolfe:2008zy} where the
$\overline{\uD 3}$-brane backreaction was studied in the
Klebanov-Tseytlin (KT) region~\cite{Klebanov:2000nc}.  However, due to
the decreased complexity of the geometry, the squashing of the
geometry in~\cite{DeWolfe:2008zy} is less dramatic than the squashing
in the near-tip region.  Likewise, the resulting non-ISD flux near the
tip is more complex than the non-ISD flux supporting the solution
of~\cite{DeWolfe:2008zy}.  We also discuss these issues in the KT
region.

Other than the consideration of $\overline{\uD 3}$-branes, the
existence of non-SUSY fluxes is interesting to consider for many
reasons.  It is well known (see for example~\cite{Camara:2003ku,
  Camara:2004jj, Grana:2003ek, Lust:2004dn}) that such non-SUSY fluxes
can give rise to soft SUSY-breaking terms in 4D effective theories.
Additionally, non-SUSY fluxes can play an important role in the
context of D-brane inflation\footnote{For recent reviews, see,
  e.g.,~\cite{Linde:2005dd, Cline:2006hu, Kallosh:2007ig,
    Burgess:2007pz, McAllister:2007bg, Baumann:2009ni}.}.  While the
deformed conifold can support certain non-SUSY fluxes (at least to the
level of approximation at which we work), we show below that in order
to have any non-ISD flux, the geometry must be squashed so that it is
no longer Calabi-Yau.

Perturbations to the KS solution appear in many other places in the
literature (and indeed most of our solutions have appeared elsewhere
though previously none had been identified as describing the presence
of $\overline{\uD 3}$-branes).  Using an alternative
parametrization~\cite{Pando_Zayas:2000sq,Papadopoulos:2000gj} of the
ansatz that we present below, the linearized equations of motion for
perturbations to the KS geometry have been written elsewhere as a
system of coupled first order equations, solutions for which can be
written formally in terms of
integrals~\cite{Borokhov:2002fm,Kuperstein:2003yt}.  Although writing
the equations of motion in this way can be convenient, we choose to
work directly with the linearized second order equations.  The second
derivative equations were also directly solved in~\cite{Apreda:2003gc,
  Schvellinger:2004am}, though we relax some of the assumptions made
in those references.  Analysis of perturbations to KS also arise in
studies of the glueball spectrum of the dual theory~\cite{Berg:2005pd,
  Berg:2006xy, Dymarsky:2008wd}.

There are several other reasons why we are interested in analyzing
non-SUSY perturbations to the near-tip region of KS. First of all,
being closest to the source of SUSY breaking, this is the region where
the supergravity fields are most affected.  Moreover, as is common in
warped compactification, the wavefunctions of non-zero modes (e.g. the
gravitino after SUSY breaking) tend to peak in the tip region.  Thus,
our perturbative solutions are useful in determining the low energy
couplings (including soft masses) in the 4D effective action involving
these infrared localized fields.  Additionally, as discussed in a
companion paper \cite{McGuirk:2009am}, the backreacted $\overline{\uD
  3}$ solution in the near tip region provides a holographic dual of
gauge mediation in a different parameter space regime from that of
\cite{Benini:2009ff}. As a result, strongly coupled messengers (and
not only weakly coupled mesonic bound states) of the hidden sector
gauge group can contribute significantly to visible sector soft terms.
Given the aforementioned applications, it is of importance for us to
consider warped geometries which are infrared smooth\footnote{By this
  we mean that, at least before the addition of 3-branes, the warp
  factor approaches a constant, or, equivalently, the (minimal
  surface) dual Wilson loop \cite{Rey:1998ik,Maldacena:1998im} has a
  finite tension.} before perturbations.  Since we are focusing on the
near tip region, our starting point is the KS solution which provides
a more accurate description at small radius than KT.  Although the KT
background correctly reproduces the cascading behavior of the field
theory, it becomes singular in the IR where the effective $\uD 3$
charge (which is dual to the scale dependent effective 't~Hooft
coupling) becomes negative and the cascade must end.  The appropriate
IR modification is the KS solution which is built on the deformed
conifold so that the solution is smooth even in the IR.

This paper is organized as follows.  In Section
\ref{sec:iib-ansatz-ks}, we discuss our solution ansatz and express
the KS solutions in accordance with this ansatz.  In Section
\ref{sec:non-susy-deformation}, we examine singular perturbations to
the warped deformed conifold and describe how we obtain the
perturbative solution corresponding to placing a
$\overline{\uD3}$-brane point source in the warped deformed conifold.
We also clarify that solutions where the internal space is unsquashed
should satisfy the ISD condition and cannot describe the backreaction
of an $\overline{\uD3}$-brane.  In Section \ref{sec:other-non-susy},
we present regular solutions which also break supersymmetry, but do
not correspond to the backreaction of a localized source. In Section
\ref{sec:comments-non-susy}, we present solutions in the KT region,
both with and without the ISD condition imposed.  We calculate the
gravitino mass in these SUSY-breaking warped backgrounds in Section
\ref{sec:gravitino-mass-} and end with some discussions in Section
\ref{sec:discussions}.  Some useful details about our conventions and
the complex coordinates of KS are relegated to the appendices.

We note that after the completion of this paper, another
preprint~\cite{Bena:2009xk} that addresses the question of adding
$\overline{\uD 3}$-branes to the geometry was made available.  Our
treatment of the $\overline{\uD 3}$-brane differs
from~\cite{Bena:2009xk} by the boundary conditions imposed in the
IR\footnote{Additionally, the equations of motion
  in~\cite{Bena:2009xk} are formally solved for all radii and would
  thus be useful for further analysis connecting the IR and UV
  regions.} as elaborated in~Sec~\ref{sec:boundary-condition}.

\newpage

%%%%%
\section{\label{sec:iib-ansatz-ks}Supergravity ansatz}
%%%%%

In this section we give the ansatz that we will use for the metric and
other fields, working in the Einstein frame of IIB supergravity.  Our
conventions are presented in Appendix~\ref{sec:notations}.  Since we
are considering perturbations to the KS
solution~\cite{Klebanov:2000hb} (see also~\cite{Herzog:2002ih}), our
ansatz will be based on that solution.  In particular, since we are
looking for perturbations that preserve the isometry of KS, we take the
metric
\begin{subequations}
  \label{geometry ansatz}
  \begin{align}
    ds^2 =& h^{-{\frac{1}{2}}}(\tau)dx_\mu^2 + h^{\frac{1}{2}}(\tau)
    d\tilde{s}_6^2,\\
    d\tilde{s}_6^2=& p(\tau) d\tau^2 + b(\tau) g_5^2 + q(\tau) (g_3^2
    + g_4^2) + s(\tau) (g_1^2 + g_2^2),
  \end{align}
\end{subequations}
where $\tau$ is the radial coordinate and where the angular one-forms
$g_{i}$ are reviewed in Appendix~\ref{sec:notations}.  This metric
ansatz includes the warped deformed conifold as a special case by a
certain choice of $p$, $b$, $q$, and $s$ presented in the next
section.  For the axiodilaton we take
\begin{subequations}
  \label{gauge ansatz}
\begin{align}
  \Phi =& \Phi(\tau) \qquad  C=0,
\end{align}
while for the fluxes,
\begin{align}
  B_2 =& \frac{g_s M \alpha'}{2}\left[f(\tau) g_1\wedge g_2 +
    k(\tau) g_3\wedge g_4\right],\notag\\
  F_3 =& \frac{M\alpha'}{2}\left[(1-F(\tau)) g_5\wedge g_3\wedge g_4 +
    F(\tau) g_5\wedge g_1\wedge g_2 \right. \notag \\
  &\phantom{\frac{M\alpha'}{2}\left[\right.}\left.+
    F'(\tau) d\tau \wedge (g_1\wedge g_3 + g_2\wedge g_4)\right],\notag\\
  F_{5}=&\left(1+\ast_{10}\right)\mathcal{F}_{5},\notag \\ {\cal F}_5 =&
  \frac{g_s M^2 \alpha'^2}{4}\ell(\tau) g_1\wedge g_2\wedge g_3\wedge
  g_4\wedge g_5.
\end{align}
\end{subequations}
These choices of fluxes respect the isometries of the deformed
conifold and satisfy the Bianchi identities
\begin{equation}
  dF_{3}=0,\qquad dH_{3}=0.
\end{equation}

%%%
\subsection{Klebanov-Strassler solution and its expansion near the
  tip}
%%%

The KS solution~\cite{Klebanov:2000hb} corresponds to placing $M$
fractional $\uD 3$-branes at a deformed confiold point (i.e. wrapping
$M$ $\uD 5$-branes around the collapsing two-cycle) and smearing these
branes over the finite ${\mathbb S}^3$.  It is recovered by the choice
\begin{align}
  \label{KS solution}
  f_{KS}(\tau)=& \frac{\tau \cosh \tau - \sinh \tau}{4 \cosh^2
    (\tau/2)}, \quad k_{KS}(\tau) = \frac{\tau \cosh \tau - \sinh
    \tau}{4 \sinh^2 (\tau/2)},\quad
  F_{KS}(\tau) = \frac{\sinh \tau -\tau}{2\sinh \tau},\notag\\
  \ell_{KS}(\tau) =& f_{KS}(1-F_{KS}) + k_{KS} F_{KS} ,\notag\\
  p_{KS}(\tau) =& b_{KS}(\tau) = \frac{\varepsilon^{4/3}}{6
    K^2(\tau)},\quad q_{KS}(\tau) = \frac{\varepsilon^{4/3}}{2}
  K(\tau)\cosh^2{\frac{\tau}{2}},\\
  s_{KS}(\tau) =& \frac{\varepsilon^{4/3}}{2}
  K(\tau)\sinh^2\frac{\tau}{2},	\notag\\
  \Phi\left(\tau\right) =& \log g_s,\quad h_{KS}(\tau) = (g_s M
  \alpha')^2 2^{\frac{2}{3}}\varepsilon^{-\frac{8}{3}}
  I (\tau),\notag\\
  K(\tau) =& \frac{\left(\sinh(2\tau) - 2\tau\right)^{1/3}}
  {2^{1/3}\sinh\tau},\quad I(\tau) = \int^{\infty}_{\tau} dx\,
  \frac{x\coth x -1}{\sinh^2 x} (\sinh 2x - 2x)^{\frac{1}{3}},
\end{align}
where $\ell_{KS}$ is determined by requiring ${\cal F}_5 = B_2 \wedge
F_3$, which ensures that the solution is regular.  However, if a
regular $\uD 3$-brane is added to the geometry, then $\ell$ must have
an additional constant part.  This introduces a $\tau^{-1}$ part to
the warp factor and the solution becomes singular.  On the field
theory side, this corresponds to the loss of confinement.

We ar[e interested in solving for perturbations to the KS background.
Since the geometry is already relatively complicated, we will consider
perturbations to the geometry as a power expansion about $\tau=0$.  It
is therefore useful to to note the series solutions for the KS
solution near the tip,
\begin{subequations}
\begin{align}
  f_{KS}&=\frac{\tau ^3}{12}-\frac{\tau ^5}{80}+\frac{17 \tau
    ^7}{10080} -\frac{31 \tau ^9}{145152}+\frac{691 \tau
    ^{11}}{26611200}+ \cdots,
  \notag\\
  k_{KS}&=\frac{\tau }{3}+\frac{\tau ^3}{180}-\frac{\tau ^5}{5040}
  +\frac{\tau ^7}{151200}-\frac{\tau ^9}{4790016}+ \cdots,\notag\\
  F_{KS}&=\frac{\tau ^2}{12}-\frac{7 \tau ^4}{720}+\frac{31 \tau
    ^6}{30240} -\frac{127 \tau ^8}{1209600}+\frac{73 \tau
    ^{10}}{6842880}+\cdots,
\end{align}
for the three-form fields, and
\begin{align}
  p_{KS}&=b_{KS}= \frac{\varepsilon ^{4/3}}{ 2^{2/3}\  3^{1/3}\ 2}
  +\frac{\varepsilon ^{4/3} \tau ^2}{ 2^{2/3}\ 3^{1/3}\ 10}
  +\frac{\varepsilon^{4/3} \tau ^4}{ 2^{2/3}\ 3^{1/3}\ 210}\notag \\
  &\hspace{10em} +\frac{\left(\frac{2}{3}\right)^{1/3} \varepsilon
    ^{4/3} \tau ^8} {606375} -\frac{19 \varepsilon ^{4/3} \tau ^{10}}
  {2^{2/3}\ 3^{1/3}\ 322481250}
  + \cdots,\notag\\
  q_{KS}&=\frac{\varepsilon ^{4/3}}{2^{2/3}\ 3^{1/3}}+
  \frac{\left(\frac{3}{2}\right)^{2/3}}{20} \varepsilon ^{4/3} \tau^2
  +\frac{17 \varepsilon ^{4/3} \tau ^4}{ 2^{2/3}\ 3^{1/3}\ 2800}\notag \\
  &\hspace{10em}+\frac{\varepsilon ^{4/3} \tau ^6}{ 2^{2/3}\ 
    3^{1/3}\ 10080} -\frac{83 \varepsilon ^{4/3}\tau ^8}{ 2^{2/3}\ 
    3^{1/3}\ 155232000}
  +\cdots, \notag\\
\intertext{\newpage}
  s_{KS}&=\frac{\varepsilon ^{4/3} \tau ^2}{4\ 2^{2/3}\ 3^{1/3}}
  -\frac{\varepsilon ^{4/3} \tau ^4}{ 2^{2/3}\ 3^{1/3}\ 240}
  +\frac{59\varepsilon ^{4/3} \tau ^6}{ 2^{2/3}\ 3^{1/3}\ 50400}\notag \\
  &\hspace{10em}-\frac{\varepsilon ^{4/3} \tau ^8}{ 2^{2/3}\ 
    3^{1/3}\ 8960} +\frac{6401 \varepsilon ^{4/3} \tau ^{10}}{ 2^{2/3}\ 
    3^{1/3}\ 558835200}
  +\cdots,\notag\\
  h_{KS}&=(g_s M \alpha')^2 2^{\frac{2}{3}} \varepsilon^{-\frac{8}{3}}
  \left[a_0 -\frac{3 \tau ^2}{ 6^{1/3}} +\frac{\tau ^4}{ 6^{1/3}\ 18}
    -\frac{37 \tau ^6}{ 6^{1/3}\ 4725} +\frac{4\, 2^{2/3}\tau ^8}{
      3^{1/3}\ 7875}+\cdots \right],
\end{align}
\end{subequations}
for the squashing functions and warp factor.
Note that each of these functions contains powers of $\tau$ with only
one parity (e.g. $h_{KS}$ contains only terms of the form
$\tau^{2k}$).  The given expansions satisfy the dilaton equation of
motion~(\ref{eq:dilatoneom}) up to $O(\tau^{9})$, the Einstein
equation~\eqref{eq:einstein} up to $O(\tau^7)$, the $H_3$
equations~(\ref{H3eom 1}) and~\eqref{H3eom 2} up to $O(\tau^8)$ and
$O(\tau^{10})$, and the $F_3$ equation~(\ref{F3eom}) up to
$O(\tau^9)$.  The leading constant of $h$ can be caluculated
numerically, $a_0 = I(0)\approx 0.71805$.

%%%%%
\section{\label{sec:non-susy-deformation}Non-SUSY deformations from
  localized sources}
%%%%%

In this section, we present perturbations to
the KS geometry that are solutions to the supergravity equations of
motion with singular behavior.  One of these solutions corresponds to
adding $\overline{\uD 3}$-branes to the tip of the geometry.  We first
argue why a singular solution is necessary to describe the point
source behavior of a $\overline{\uD 3}$.  We then discuss two
solutions, the first in which the internal space remains the deformed
conifold, and the second in which the geometry is ``squashed'' away
from this geometry.  We also match the parameters in the latter
solution to the tension and charge of the $3$-branes.
 
%%%
\subsection{Effect of point sources\label{sec:effect-point-source}}
%%%

For simplicity, we seek solutions that retain the isometry of the KS
solution.  Because the ${\mathbb S}^3$ remains finite as $\tau=0$, in
general placing a point source into the internal geometry will break
the angular isometry even at $\tau=0$.  Therefore, in order to retain
the KS isometry, these point sources must be smeared over the finite
${\mathbb S}^3$.  Alternatively, we can consider a collection of point
sources that to good approximation are uniformly distributed over the
${\mathbb S}^3$.

The effect of such a localized source on the geometry can be estimated
by considering the Green's function in the unperturbed background.
Using the metric~\eqref{geometry ansatz}, an $S$-wave solution
(i.e. dependent only on $\tau$) to Laplace's equation
$\nabla^{2}H=0$ can be written with integration constants
$\mathcal{P}$ and $\phi$
\begin{equation}
  \label{eq:greensfunction}
  H={\cal P} - 
  \int^\infty_\tau dx\, \frac{\phi}{q(x) s(x)} \sqrt{\frac{p(x)}{b(x)}},
\end{equation}

If the warp factor $h$ is obtained by solving the Killing spinor
equations~\eqref{susy variations} (more precisely
(\ref{eq:warpfactorsusy})), then the form of $h$ is similar to that of
the Green's function,
\begin{equation}
  h= {\cal P} + \frac{(g_s M \alpha')^2}{4}
  \int_\tau^{\infty} d x \frac{\ell(x)}{q(x) s(x)}
  \sqrt{\frac{p(x)}{b(x)}}.
  \label{supersymmetric warp factor}
\end{equation}
A localized source of $\uD 3$ charge gives a constant piece to $\ell$,
and this constant piece indeed generates a Green's function in $h$.
The ansatz~\eqref{geometry ansatz} admits solutions that are either
asymptotically flat or AdS (up to possible log corrections) and
setting $\mathcal{P}=0$ corresponds to demanding the latter.

The integral~\eqref{eq:greensfunction} cannot be performed exactly for
the KS solution.  However, using the expansion about $\tau=0$ given
above, one can write for small $\tau$,
\begin{equation}
  H = {\cal P} + 2^{\frac{10}{3}}3^{\frac{2}{3}}\,\phi\,
  \varepsilon^{-\frac{8}{3}}
  \left(-\frac{1}{\tau} - \frac{2\tau}{15} + \frac{\tau^3}{315} +
    \cdots \right).
  \label{Green function near tip of the deformed conifold}
\end{equation}
Thus if we include a point source at the tip that respects the same
supersymmetry as KS, then the geometry should become singular as
$\tau\to 0$.  In particular, the warp factor will depend as
$\tau^{-1}$.

Dropping the constant term, in the large radius region the Green's
function takes the form (in terms of
$r^{2}=2^{5/3}3\ve^{4/3}e^{2\tau/3}$)
\begin{equation}
  H = - 2^{1/3} 24\, \varepsilon^{-{\frac{8}{3}}}\,\phi\, 
  e^{- {\frac{4\tau}{3}}} + \cdots
  = - \frac{27\, \phi}{r^4}+ \cdots.
  \label{Green function far tip}
\end{equation}

If the object that is added to geometry breaks supersymmetry, then it
needs not perturb the warp factor by simply adding a Green's function
piece.  Indeed, in the large radius region, where a Green's function
behaves as $r^{-4}$, adding $\uD 3$-$\overline{\uD 3}$ pairs perturbs
the warp factor by $r^{-8}$~\cite{DeWolfe:2008zy,Brax:2000cf} (though
there are also log corrections).  Heuristically, the presence of the
non-supersymmetric source adds a perturbation that scales as $\delta
h\sim h_{0}H$ where $h_{0}$ is the unperturbed warp factor (if the
charge of the non-BPS source is non-vanishing, then there will be a
Green's function contribution as well).  Since the KS warp factor
approaches a constant, this suggests that even for non-supersymmetric
sources, we should look for perturbations to the warp factor that
behave as $\tau^{-1}$.  This argument is only very heuristic, though
we are able to check using boundary conditions that this behavior is
indeed correct.

%%%
\subsection{\label{sec:pert-solut-deform}Unsquashed singular
  perturbations to KS}
%%%

We consider perturbing the KS solution by taking the
ansatz~\eqref{geometry ansatz},~\eqref{gauge ansatz} and writing
\begin{align}
  f=& f_{KS} + f_{p}(\tau),\quad k=k_{KS}+ k_{p}(\tau),\quad
  F=F_{KS}+ F_{p}(\tau),\quad \ell = f(1-F) + k F,\notag\\
  \Phi =& \log g_s + \Phi_{p}(\tau),\quad h = h_{KS} + h_p(\tau),\notag \\
  p=&p_{KS},\quad b=b_{KS},\quad q=q_{KS},\quad s=s_{KS},
  \label{non-squashing perturbative ansatz}
\end{align}
where the KS solution is~\eqref{KS solution} and the subscript $p$
indicates a perturbation to KS.  Such an ansatz changes the warp
factor and the fluxes, but leaves the internal unwarped geometry as
the deformed conifold.  We do not attempt to solve for the
perturbations exactly, but write them as a power series about
$\tau=0$.  We then solve the equations of motion~\eqref{eom} to first
order in the perturbations and order-by-order in $\tau$.  To linear
order in perturbations the coefficients for the odd powers of $\tau$
in $h_{p}$ decouple from the coefficients for the even powers.  As
argued above, to capture the behavior of a point source, the warp
factor ought to behave as $\tau^{-1}$, implying that we should focus
on the odd powers in $\tau$ in $h_{p}$.  We find the solution
\begin{align}
  \Phi_p=& \phi\left(\frac{1}{\tau} + \frac{2\tau}{15} -
    \frac{\tau^3}{315} +
    \frac{2 \tau^5}{23625}\right), \notag\\
  F_p=&\phi\left(- \frac{1}{2} - \frac{23 \tau^3}{720} +
    \frac{\tau^5}{1400}\right) + {\cal U}\left(\frac{1}{\tau} -
    \frac{\tau}{6}+
    \frac{7\tau^3}{360} - \frac{31 \tau^5}{15120}\right), \notag\\
  f_p=& \phi\left(\frac{23}{12} + \frac{3\tau^2}{16} -
    \frac{\tau^4}{80} + \frac{61 \tau^6}{26880} \right) + {\cal U}
  \left(- \frac{13}{6} - \frac{\tau^2}{8}+ \frac{\tau^4}{48}-
    \frac{17\tau^6}{5760}\right) +
  \frac{{\cal H}}{6},\notag\\
  k_p =& \phi\left(\frac{1}{\tau^2} + \frac{9}{4} - \frac{3
      \tau^2}{80} - \frac{113 \tau^4}{25200}\right) + {\cal
    U}\left(-\frac{2}{\tau^2} -\frac{5}{2} - \frac{\tau^2}{120} +
    \frac{\tau^4}{3024}\right) +
  \frac{{\cal H}}{6},\notag\\
  h_p =& \frac{(g_s M \alpha')^2 2^{2/ 3}
    \varepsilon^{-{8/3}}}{6^{1/3}} \left[\phi\left(\frac{11}{\tau} +
      \frac{206 \tau^3}{1575} - \frac{487 \tau^5}{23625}\right)+ {\cal
      U}\left(- \frac{12}{\tau} - \frac{4 \tau^3}{25} +
      \frac{208 \tau^5}{7875}\right) \right. \notag\\
  & \hspace{10em} \left. + {\cal H}\left(\frac{1}{\tau} + \frac{2
        \tau}{15} - \frac{\tau^3}{315} + \frac{2 \tau^5}{23625}\right)
  \right].
  \label{singular simplest solution}
\end{align}
This solution is valid to linear order in the parameters $\phi$,
${\cal U}$, and $\mathcal{H}$.  It can be extended to higher order in
$\tau$ by expressing the higher order coefficients in terms of $\phi$,
$\mathcal{U}$, and $\mathcal{H}$ so that no additional parameters need
to be introduced.  These perturbations satisfy the dilaton
equation~\eqref{eq:dilatoneom} up to $O(\tau^3)$, the Einstein
equation~\eqref{eq:einstein} up to $O(\tau^3)$, and the gauge
equations~(\ref{H3eom 1}) up to $O(\tau^4)$,~(\ref{H3eom 2}) up to
$O(\tau^{4})$, and~(\ref{F3eom}) up to $O(\tau^{3})$.

It is worth noting that even if we allow a perturbation to $b(\tau)$,
which describes limited squashing of the internal space (more general
squashing is considered below), the solution~\eqref{singular simplest
  solution} does not change and $b$ remains unperturbed ($b_p =0$).
The squashing of this direction was considered
in~\cite{DeWolfe:2008zy} to obtain a non-SUSY deformation of KT space,
but in the KS region, there is no solution in which only this
direction is squashed.

To this order in the perturbations and in $\tau$, the
solution~\eqref{singular simplest solution} respects the ISD
condition~(\ref{ISD & IASD condition}) of the $3$-form flux as well as
the first derivative SUSY condition for the warp
factor~(\ref{eq:warpfactorsusy}), even though the solution follows
from solving second derivative equations.  However, we expect (and
indeed we have checked to several higher orders in $\tau$) that the
flux remains ISD to all orders in $\tau$ since the dilaton takes the
form of a Green's function~\eqref{Green function near tip of the
  deformed conifold}.  If the flux had an IASD component as well, then
in general the fluxes would provide a potential for the dilaton and
$\Phi$ would no longer satisfy $\nabla^{2}\Phi=0$.  Indeed, since
$\Phi$ does have the same form
as~\eqref{eq:greensfunction}, we can identify $\phi$ as
corresponding to some point source for the dilaton smeared over the
finite ${\mathbb S}^3$ at $\tau=0$.

Some non-SUSY perturbations to the KS solutions, found by solving the
first order differential equations given in~\cite{Pando_Zayas:2000sq},
were analyzed in~\cite{Borokhov:2002fm,Kuperstein:2003yt}.  For
$\phi=0$ (i.e. constant dilaton), the solution~\eqref{singular
  simplest solution} is a small $\tau$ expansion of the exact solution
appearing in~\cite{Kuperstein:2003yt}, the flux part of which is
\begin{align}
  F\bigl(\tau\bigr) =& \frac{1}{2}\left(1-\frac{\tau}{\sinh
      \tau}\right)+ \frac{{\cal U}}{\sinh \tau} + \frac{5{\cal
      T}}{32}\left(\cosh \tau - \frac{\tau}{\sinh \tau}
  \right),\notag\\
  f(\tau) =& \frac{\tau \cosh \tau - \sinh \tau}{4 \cosh^2 (\tau/2)} -
  \frac{{\cal U}}{6\cosh^2 (\tau/2)}\left(5 + 8 \cosh \tau \right) +
  \frac{{\cal H}}{6}\notag\\
  &+\frac{5{\cal T}}{128 \cosh^2 (\tau/2)} \left(2\tau + 4 \tau \cosh
    \tau -4 \sinh\tau - \sinh 2\tau \right),
  \notag\\
  k(\tau) =& \frac{\tau \cosh \tau - \sinh \tau}{4 \sinh^2 (\tau/2)}-
  \frac{{\cal U}}{6 \sinh^2 (\tau/2)}\left(-5 + 8 \cosh \tau \right) +
  \frac{{\cal H}}{6}\notag\\
  &+\frac{5{\cal T}}{128\sinh^2 (\tau/2)} \left(-2\tau + 4 \tau \cosh
    \tau -4 \sinh\tau + \sinh 2\tau \right).
  \label{Exact ISD solution with constant dilaton}
\end{align}
The solution is singular for non-vanishing ${\cal U}$ or $\mathcal{H}$
which are essentially the same parameters that appear
in~(\ref{singular simplest solution}), though~\eqref{Exact ISD
  solution with constant dilaton} is an exact solution of~\eqref{eom}
to all orders in $\mathcal{U}$, $\mathcal{H}$, and $\mathcal{T}$.  The
remaining parameter ${\cal T}$ appears in another
solution~(\ref{simplest regular solution}), and of the parameters
of~\eqref{Exact ISD solution with constant dilaton}, only a
non-vanishing $\mathcal{T}$ leads to supersymmetry breaking.  The
additional parameter $\phi$ appearing in~\eqref{singular simplest
  solution} comes from relaxing the condition that the dilaton $\Phi$
is constant.  Note also that the parameter ${\cal H}$ is related to
the gauge symmetry $B_2 \rightarrow B_2 + d \Lambda_1$.

To check if supersymmetry is preserved, we consider the SUSY
variations of the gravitino and dilatino~\eqref{susy variations},
taking into account the non-trivial dilaton profile.  Since $G_{3}$ is
ISD, the last term of the dilatino variation~\eqref{eq:dilatinovar}
vanishes.  However, the terms involving the derivative of the dilaton
do not.  Indeed for small $\tau$,
\begin{equation}
  \delta \lambda \sim - i\frac{3^{1/6}\phi}
  {2^{1/3} a_0^{1/4} (g_s M \alpha')^{1/2} \tau^2}
  \hat{\Gamma}^\tau \epsilon + \cdots,
\end{equation}
where $\hat{\Gamma}$ indicates an unwarped $\Gamma$-matrix.  Since
this variation is non-vanishing, the solution~\eqref{singular simplest
  solution} breaks supersymmetry.

The variation for the gravitino is also non-vanishing since the
solution includes a $\left(0,3\right)$ part of $G_{3}$.
From~\eqref{(0,3) and (3,0)-forms}, we see that for the
solution~\eqref{singular simplest solution},
\begin{equation}
  G_3^{(0,3)} = \phi \left( \frac{1}{3 \tau^3} + \frac{1}{15 \tau} -
    \frac{86 \tau}{1575} + \cdots \right)
  (z_i d\bar{z}_i)\wedge
  (\epsilon_{ijkl} z_i \bar{z}_j d\bar{z}_k \wedge d\bar{z}_l).
\end{equation}
As shown in Sec.~\ref{sec:solut-with-regul}, the exact
solution~\eqref{Exact ISD solution with constant dilaton}, for which
$\phi=0$, has an additional contribution to the $\left(0,3\right)$
part from $\mathcal{T}$.  For the perturbation~\eqref{singular
  simplest solution}, the $\left(3,0\right)$ and
$\left(1,2\right)$-parts vanish, which is consistent with the fact
that the ISD condition allows only for $\left(2,1\right)$ and
$\left(0,3\right)$ components.

Both the variation of the dilatino and gravitino involve only $\phi$.
Therefore, even though the singular behavior seems to imply that
${\cal U}$ and $\mathcal{H}$ can be associated with a point source,
they do not break supersymmetry (though~\eqref{Exact ISD solution with
  constant dilaton} breaks supersymmetry for non-vanishing
$\mathcal{T}$) and only $\phi$ is a possible candidate to describe the
presence of a localized SUSY-breaking source.  The parameter $\phi$
characterizes a localized source for the dilaton and therefore cannot
correspond to the presence of $\overline{\uD 3}$-branes since
$\overline{\uD 3}$s do not directly couple to the dilaton.
Furthermore, it was shown in~\cite{DeWolfe:2008zy} that
an~$\overline{\uD 3}$ squashes the geometry so that it is no longer
conformally Calabi-Yau.  Extrapolating this result to short distances,
the source associated with $\phi$, which does not squash the geometry,
should therefore not be identified with an $\overline{\uD 3}$-brane.
Indeed, this mode is the small radius analogue of the $r^{-4}$ mode
for the dilaton that appeared in~\cite{DeWolfe:2008zy} (as well as the
flat space analysis of non-BPS branes in~\cite{Brax:2000cf}) which
could be turned off independently of the existence of $\overline{\uD
  3}$-branes as it does not contribute to the total mass of the
solution.

Note that this solution possesses a curvature singularity at $\tau=0$;
at small $\tau$ the Ricci scalar behaves as
\begin{equation}
  R = \frac{-45 \, 2^{2/3} {\cal H} + (45 \ 2^{2/3} - 3^{1/3} a_0)
    (12 {\cal U} - 11 \phi)}{30 3^{1/3} a_0^{5/2} g_s M \alpha' \tau}
  + \cdots.
\end{equation}
The presence of the curvature singularity indicates a breakdown of the
supergravity approximation, and so our solution is only expected to be
valid for $1/\left(g_{s}M\alpha'\right)\ll \tau < 1$ where the upper
bound coming from the fact that we are performing a small $\tau$
expansion and the lower bound comes from assuming that $R$ is small in
string units.

%%%
\subsection{\label{sec:squashedsingularks}Squashed singular
  perturbations to KS}
%%%

We can generalize by considering solutions that ``squash'' the
internal geometry so that the unwarped geometry is no longer that of
the deformed conifold.  At large distances where the DKM
solution~\cite{DeWolfe:2008zy} is valid, the only non-trivial
squashing that occurs due to the presence of a $\overline{\uD
  3}$-brane is in the direction on which the $U\left(1\right)$
isometry acts\footnote{In actuality, this $U\left(1\right)$ isometry
  is broken to a discrete subgroup by the fluxes and deformation of
  the conifold singularity.}.  However, as discussed in the previous
section, at small radius the equations of motion do not admit a
solution in which the only squashing is in this direction.  Thus, we
consider 
\begin{align}
  \Phi =& \log g_s + \Phi_{p}(\tau),\quad
  h = h_{KS} + h_p(\tau),\notag\\
  f=& f_{KS} + f_{p}(\tau),\quad k=k_{KS} + k_{p}(\tau),\quad F=F_{KS}
  + F_{p}(\tau),\quad
  \ell = f(1-F) + k F, \notag\\
  b=&b_{KS}(1 + b_p(\tau)),\quad q = q_{KS}(1+ q_p(\tau)), \quad s=
  s_{KS}(1 + s_p(\tau)), \quad p=p_{KS},
  \label{squashing perturbative ansatz}
\end{align}
where we have used the freedom to redefine $\tau$ to keep $p$
unperturbed but have allowed $b$, $q$, and $s$ to be general so that
the ansatz is the most general ansatz consistent with the isometry of
KS.  This more general ansatz will allow $G_{3}$ to have both ISD and
IASD components. We are again interested in describing the effect of a
localized source and since the even and odd powers of $\tau$ in the
warp factor decouple from each other, we focus on odd powers of $\tau$
in $h_{p}$.  We find a power series solution to~\eqref{eom} where the
dilaton obtains a non-trivial profile that is regular at small $\tau$
\begin{subequations}
  \label{singular general solution}
\begin{equation}
  \Phi_p = {\cal S} \tau +\mathcal{Y} \tau ^3.
\end{equation}
However, the squashing functions for the solution are singular
\begin{align}
  b_p =& {\cal S} \left(\frac{7}{\tau }- \frac{3293 \tau
      ^3}{3150}\right) + \mathcal{Y} \left(\frac{70}{\tau }-\frac{404
      \tau ^3}{45}\right)+
  {\cal B}\left(\tau -\frac{43 \tau ^3}{210}\right),\notag\\
  q_p =&{\cal S} \left(\frac{7}{4 \tau }+ \frac{103\tau
    }{48}-\frac{44129 \tau ^3}{100800}\right)+ \mathcal{Y}
  \left(\frac{35}{2 \tau }+\frac{70 \tau }{3}-\frac{1673 \tau ^3}
    {360}\right)\notag \\
  &+{\cal B} \left(\frac{3 \tau }{4}-\frac{71 \tau
      ^3}{560}\right),
  \notag\\
  s_p = &{\cal S} \left(\frac{73}{3 \tau }-\frac{253 \tau
    }{720}+\frac{29999 \tau ^3}{60480}\right)+\mathcal{Y} \left(
    \frac{1085}{6 \tau }-\frac{56 \tau }{9}+\frac{1049 \tau
      ^3}{216}\right)
  \notag \\
  &+{\cal B} \left(\frac{5}{\tau }-\frac{7 \tau }{12}+
    \frac{529 \tau ^3}{5040}\right).
\end{align}
Similarly the fluxes are
\begin{align}
  F_p =&{\cal S} \left(\frac{3193}{84
      \tau}+\frac{312}{35 \tau} 6^{1/3} a_0+
    \left(-\frac{5959}{1008}-\frac{299 \ 3^{1/3} a_0}{70 \
        2^{2/3}}\right)
    \tau \right) \notag\\
  &+\mathcal{Y} \left({\frac{760}{3\tau}+\frac{72 \ 6^{1/3}
        a_0}{\tau}}+ \left(-\frac{1555}{36}-\frac{69 \ 3^{1/3} a_0}{2
        \ 2^{2/3}}\right) \tau
  \right)\notag\\
  &+{\cal B} \left(\frac{50}{7 \tau }-\frac{65 \tau }{84}\right),\notag\\
\intertext{\newpage}
  f_p = & {\cal S} \left(-\frac{863221}{2520}-\frac{101247\ 3^{1/3}
      a_0}{700 \ 2^{2/3}}+\frac{13 \ 3^{2/3} a_0^2}{5 \
      2^{1/3}}+\left(-\frac{265}{112}-\frac{321 \ 3^{1/3} a_0}{112 \
        2^{2/3}}\right)
    \tau ^2 \right.\notag\\
  &\left. \quad +\left(\frac{17209}{40320}+\frac{1037\ 3^{1/3}
        a_0}{5600\ 2^{2/3}}\right) \tau
    ^4+\left(\frac{143221}{2419200}-\frac{653 a_0}{6400
        6^{2/3}}\right) \tau ^6\right)\notag\\
  &+\mathcal{Y} \left(-\frac{81913}{36}-\frac{5333\ 3^{1/3} a_0}{5\
      2^{2/3}}+\frac{21 \ 3^{2/3} a_0^2}{2^{1/3}}+
    \left(-\frac{125}{8}-\frac{165 \ 3^{1/3} a_0}{8\ 2^{2/3}}\right)
    \tau ^2
  \right.\notag\\
  &\left. \quad +\left(\frac{491}{288}+\frac{43\ 3^{1/3} a_0}{40\
        2^{2/3}}\right) \tau ^4+\left(\frac{3881}{4320}-
      \frac{413 a_0}{640\  6^{2/3}}\right) \tau ^6\right)\notag\\
  & + {\cal B} \left(-\frac{3155}{84}-\frac{15 \tau ^2}{56}-
    \frac{47 \tau ^4}{672}+\frac{307 \tau ^6}{10080}\right),\notag\\
  k_p =&{\cal S} \left(-\frac{3193}{42\tau^2 }-\frac{513 \ 3^{1/3}
      a_0}{35 \ 2^{2/3} \tau^2} -\frac{37454}{105}-\frac{104777\ 
      3^{1/3} a_0}{700 \ 2^{2/3}}+\frac{13 \ 3^{2/3} a_0^2}{5 \
      2^{1/3}}\right.
  \notag\\
  &\quad \left. +\left(-\frac{11617}{5040}+\frac{1649\
        3^{1/3} a_0}{2800 \ 2^{2/3}}\right) \tau^2\right)\notag\\
  & +\mathcal{Y} \left({-\frac{1520}{3\tau^2}- \frac{39\ 6^{1/3}
        a_0}{\tau^2}}-\frac{28621}{12}-\frac{5503 \ 3^{1/3} a_0}
    {5 \ 2^{2/3}}+\frac{21 \ 3^{2/3} a_0^2}{2^{1/3}} \right.\notag\\
  &\left. \quad +\left(-\frac{1321}{72}+\frac{197 \ 3^{1/3} a_0}
      {40 \ 2^{2/3}}\right) \tau ^2\right)\notag\\
  &+{\cal B} \left(-\frac{100}{7 \tau ^2}-\frac{1095}{28}-
    \frac{17 \tau ^2}{168}\right).\notag\\
\end{align}
The warp factor resulting from the fluxes exhibits the desired singular
behavior
\begin{align}
  h_p = &(g_s M \alpha')^2
  2^{\frac{2}{3}} \varepsilon^{-\frac{8}{3}} \left[ {\cal S}
    \left(-\frac{105907\ 2^{2/3}}{105 \ 3^{1/3} \tau}-\frac{146913
        a_0}{350 \tau}+ \left(-\frac{182561}{900 \ 6^{1/3}}-\frac{5413
          a_0}{125}\right)
      \tau \right) \right. \notag\\
  & \left. +\mathcal{Y} \left(-\frac{80393} {6\ 6^{1/3}
        \tau}-\frac{30753 a_0}{10\tau}+\left(-\frac{29414 \ 2^{2/3}}
        {45 \ 3^{1/3}}-\frac{7896 a_0}{25}\right) \tau \right)\right. \notag\\
  & \left. +{\cal B} \left(-\frac{3055}{14 \ 6^{1/3} \tau }-\frac{29 \
        2^{2/3} \tau }{3 \ 3^{1/3}}\right) \right].
\end{align}
\end{subequations}
Perturbations that respect the ISD condition and were presented in the
previous section have been omitted.  Again, ${\cal S}$, $\mathcal{Y}$,
and $\mathcal{B}$ are treated as perturbations and so the solution is
valid to linear order in these parameters and can be extended to
higher order in $\tau$ without introducing any new independent
parameters.

Since the dilaton does not exhibit a $\tau^{-1}$ behavior, the
nontrivial profile cannot be interpreted as resulting from a localized
source for the dilaton.  Instead it comes from the lift of the vacuum
energy due to the presence of both ISD and IASD components of
$G_{3}$~\eqref{eq:dilatoneom},
\begin{equation}
  H_3^2 - e^{2\Phi} F_3^2 =  \frac{48 \,6^{1/3}{\cal S}}
  {\sqrt{a_0} g_s^3 M \alpha' \tau} +
  \frac{40 {\cal S} - 16 \ 6^{1/3} a_0 ({\cal S} - 90 {\cal Y}) \tau} 
  {5 a_0^{3/2} g_s^3 M \alpha'}+ \cdots.
\end{equation}
The non-vanishing potential for the dilaton implies the existence of
an IASD component since, for $C=0$,
\begin{equation}
  \nabla^{2}\Phi=-\frac{g_{s}e^{-\Phi}}{2\times 3!}
  \biggl[H_{3}^{2}-e^{2\Phi}F_{3}^{2}\biggr]\propto
  \mathop{\rm Re}\bigl(G^{+}_{mnp}G^{-mnp}\bigr)
  \label{eq:dilatonpotential}
\end{equation}
where $G^{\pm}=iG_{3}\pm\tilde{\ast}_{6}G_{3}$.
One can also see directly from~\eqref{ISD & IASD condition} that
$G_{3}$ is no longer purely ISD.  The parameters controlling the
deviation from the ISD condition~\eqref{ISD & IASD condition} are
${\cal S}$ and ${\cal Y}$.  Both of these are included in the
squashing functions, implying that the squashing of the deformed
conifold is needed to have non-vanishing $\left(3,0\right)$ or
$\left(1,2\right)$ components of $G_3$.

As was the case for~\eqref{singular simplest solution}, the geometry
exhibits a curvature singularity at $\tau=0$.  Indeed at small $\tau$
the Ricci scalar behaves as
\begin{equation}
  \label{eq:curvature}
  R\sim\frac{{\cal O}(\mathcal{B}, \mathcal{S}, \mathcal{Y})}
  {g_{s}M\alpha'\tau},
\end{equation}
where we have omitted numerical coefficients since the essential
behavior is $\tau^{-1}$.  This singularity implies that the solution
is valid only for $\mathcal{S}/(g_s M \alpha') \ll \tau < 1$.

The solution~\eqref{singular general solution} breaks supersymmetry,
squashes the geometry, and introduces an IASD component of the flux.
All of these properties are also shared by the DKM
solution~\cite{DeWolfe:2008zy} which describes the large radius
influence of $\uD 3$-$\overline{\uD 3}$ pairs at the conifold point
(though the DKM solution contains squashing in fewer directions than
this solution).  The perturbations to the KT geometry in the DKM
solution behave as $r^{-4}$ and $r^{-4}\log r$ compared to the KT
geometry itself (e.g. the KT warp factor included $r^{-4}$ and
$r^{-4}\log r$ while the perturbations behaved as $r^{-8}$ and
$r^{-8}\log r$).  Similarly, the solution~\eqref{singular general
  solution} involves perturbations that behave as $\tau^{-1}$ relative
to the KS solution~\eqref{KS solution}.  We note that $\tau^{-1}$ and
$r^{-4}$ are the small and large radius expansions of the Green's
function~\eqref{eq:greensfunction}.  This, together with the shared
properties mentioned above, is a hint that the
solution~\eqref{singular general solution} may describe the
backreaction of $\overline{\uD 3}$-branes.  We can confirm that this
is the case by checking boundary conditions.

%%%
\subsection{\label{sec:boundary-condition}Boundary conditions}
%%%

We now seek to match the parameters of this solution to the tension of
the $\overline{\uD 3}$-branes that are localized at $\tau=0$.  Since
the solution is singular at $\tau=0$, we expect the solution to be
modified by undetermined stringy corrections at distances
$\tau\lesssim {\cal S}/\left(g_{s}M\alpha'\right)$.  We will therefore
not try to obtain the coefficients exactly.

Following from the behavior of the Green's function~\eqref{Green
  function near tip of the deformed conifold}, the
$O\left(1/\tau\right)$ behavior of the warp factor is tied to the
existence of a localized source of tension.  Indeed if there is a
collection of $\uD 3$-branes and $\overline{\uD 3}$-branes located at
$\tau=0$ and angular positions $\Omega_{i}$, then they contribute to
the stress-energy tensor as
\begin{equation}
  T_{\mu\nu}^{\mathrm{loc}}
  =-\kappa_{10}^{2}T_{3}\frac{\delta\left(\tau\right)}
  {\sqrt{p_{KS}b_{KS}}s_{KS}q_{KS}}
  \sum_{i}\frac{\delta^{5}\left(\Omega-\Omega_{i}\right)}
  {\sqrt{\tilde{g}_{5}}}
  \eta_{\mu\nu},
\end{equation}
where $\tilde{g}_{5}$ is the angular part of the determinant of the
unwarped metric and the other components of $T_{MN}^{\mathrm{loc}}$
vanish.  We make the approximation that there are enough $3$-branes
that we can treat them as uniformly smeared over the finite ${\mathbb
  S}^3$ at the tip.  Then integrating over the ${\mathbb S}^3$ gives
\begin{equation}
  \int_{{\mathbb{S}}^{3}}d\mathrm{vol}_{\mathbb{S}^{3}}\, T_{\mu\nu}^{\mathrm{loc}}
  =-\kappa_{10}^{2}T_{3}\left(N_{\uD 3}+N_{\overline{\uD 3}}\right)
  \frac{\delta\left(\tau\right)}
  {\sqrt{p_{KS}b_{KS}}s_{KS}q_{KS}}
  \frac{\delta^{2}\left(\Omega\right)}
  {\sqrt{\tilde{g}_{2}}}
  \eta_{\mu\nu},
\end{equation}
where $\delta^{2}\left(\Omega\right)$ fixes the angular position on
the vanishing 2-cycle, $\tilde{g}_{2}$ is the unwarped metric for that
2-cycle, and $N_{\uD 3}$ and $N_{\overline{\uD 3}}$ are the numbers of
$\uD 3$ and $\overline{\uD 3}$-branes added to the tip.  This
localized source of tension should cause a $1/\tau$ behavior in the
warp factor.  Tracing over the Einstein equation in the presence of
the localized source, we have
\begin{equation}
  -\frac{1}{4\tau^{2}}\partial_{\tau}\left(\tau^{2}\partial_{\tau}h_{p}\right)\sim
  \frac{1}{2}\kappa_{10}^{2}T_{3}\left(N_{\uD 3}+N_{\overline{\uD 3}}\right)
  \delta\left(\tau\right)
  \sqrt{\frac{p_{KS}}{b_{KS}}}
  \frac{1}{q_{KS}s_{KS}}
  \frac{1}{\vol_{2}},
\end{equation}
where we have integrated over the angular directions and defined
$\vol_{2}=\int d^{2}x\sqrt{\tilde{g}_{2}}$.  Integrating over $\tau$,
we find that near the tip of the deformed conifold,
\begin{equation}
  \label{eq:hpbc}
  h_p \sim \frac{\left(N_{\uD 3}+N_{\overline{\uD 3}}\right)
    \kappa_{10}^{2}T_{3}\ve^{-8/3}}{\vol_{2}}\frac{1}{\tau},
\end{equation}
where $h_{KS}=h_{0}+\mathcal{O}\left(\tau^{2}\right)$.  That is, the
$\tau^{-1}$ coefficient in the warp factor is proportional to the
total tension of the $3$-branes added to the tip.
Using~\eqref{singular general solution}, we can use this relation to
match the parameters to this tension.

Similarly, one can match to the total charge added to $\tau=0$ by
considering the constant part of $\ell$,
\begin{equation}
  \ell\left(\tau=0\right)\propto N_{\uD 3}-N_{\overline{\uD 3}}.
\end{equation}
Since the solution~\eqref{singular general solution} involves ratios
of relatively large numbers, we omit the detailed form of this
expression, but by some choice of parameters, we can take the solution
to correspond to adding negative charge.  Thus, for some choice of
$\mathcal{S}$, $\mathcal{B}$, and $\mathcal{Y}$ (as discussed below,
one must additionally include the $\mathcal{U}$-mode
of~\eqref{singular simplest solution}), the solution~\eqref{singular
  general solution} corresponds to adding $\overline{\uD 3}$-branes to
the tip of KS.  Another choice of parameters allows us to describe the
influence of $\uD 3$-$\overline{\uD 3}$ pairs which adds tension, but
no net charge to the solution and so is the small radius analogue of
the solution presented in~\cite{DeWolfe:2008zy}.

Alternatively, we might match the parameters to the tension of the
$3$-branes by calculating the analogue of the ADM mass.  For spaces
which do not necessarily asymptote to either flat space or AdS, a
generalization of the ADM mass was given in~\cite{Hawking:1995fd}.
However, this is applicable only at large distances (and indeed was
used in the large radius solution~\cite{DeWolfe:2008zy}) while the
solution~\eqref{squashing perturbative ansatz} is valid for small
$\tau$.  Although analogues of the ADM mass exist for arbitrary
surfaces, and not just those at
infinity~\cite{Hawking:1968qt,Hayward:1993ph}, it is more efficient to
match to the localized tension discussed above.

The behavior of the $3$-form fluxes in~\eqref{singular general
  solution} gives rise to divergent energy densities $H_{3}^{2}$ and
$F_{3}^{2}$.  In particular, the leading order behavior
$F_{p}\sim\mathcal{S}\tau^{-1}$ (for the remainder of the section,
$\mathcal{S}$ will be used as short hand for linear combinations of
$\mathcal{S}$, $\mathcal{B}$, and $\mathcal{Y}$) leads to
$F_{3}^{2}\sim\mathcal{S}^{2}\tau^{-6}$.  The contribution to the
action then diverges since $\sqrt{g}F_{3}^{2}\sim
\mathcal{S}^{2}\tau^{-4}$.  Similarly, the $\tau^{-2}$ behavior of
$k_{p}-f_{p}$ gives $H_{3}^{2}\sim\mathcal{S}^{2}\tau^{-4}$ which also
gives a divergent action.  Since the $\overline{\uD 3}$-branes do not
directly source these fields, one should impose that these very
singular behaviors should be absent from the solution describing the
backreaction of $\overline{\uD 3}$-branes.  Since two of
$\mathcal{S}$, $\mathcal{B}$, and $\mathcal{Y}$ are fixed by matching
to the tension and charge of the $3$-branes, there is not enough
freedom to cancel both of these divergences using just the modes
in~\eqref{singular general solution}.  However, these divergences can
be cancelled by additionally including the $\mathcal{U}$-modes given
in~\eqref{singular simplest solution}.  Imposing this additional
condition on $\mathcal{S}$, $\mathcal{B}$, $\mathcal{Y}$, and
$\mathcal{U}$ gives the leading order behavior
\begin{equation}
  F_{p}\sim \mathcal{S}\tau,\quad k_{p}-f_{p}\sim \mathcal{S}\tau^{0}.
\end{equation}
From these,
\begin{equation}
  F_{3}^{2}\sim H_{3}^{2}\sim \frac{\mathcal{S}^{2}}{\tau^{2}}.
\end{equation}
That is, even after imposing that the most singular parts of the
$3$-form flux vanish, the energy densities $H_{3}^{2}$ and $F_{3}^{2}$
are divergent.  Furthermore, these divergences cannot be removed by
including any of the other modes discussed here without setting all of
these constants to be zero.  However, these do not lead to a divergent
action since $\sqrt{g}F_{3}^{2}\sim\sqrt{g}H_{3}^{2}\sim\tau^{0}$.

The fact that $F_{3}^{2}$ and $H_{3}^{2}$ are divergent may be at
first be surprising since the $\overline{\uD 3}$s do not directly
couple to the $3$-form flux and thus the singularities in $H_{3}^{2}$
and $F_{3}^{2}$ have no obvious physical interpretation\footnote{For
  example, it was argued in~\cite{Bena:2009xk} that the resulting
  $H_{3}$ has the wrong orientation and dependence on the
  $\overline{\uD 3}$-charge to be due to the NS5-branes that were
  described in~\cite{Kachru:2002gs}.}.  Here, however, we suggest that
such singular behavior might have been anticipated from the equations
of motion and the boundary conditions.  Indeed, the coupling between
the $3$-form and $5$-form flux can be written as (see for
example~\cite{Giddings:2001yu})
\begin{equation}
  \label{eq:GKPequation}
  d\Lambda+\frac{i}{\mathrm{Im}\left(\tau\right)}
  d\tau\wedge\bigl(\Lambda+\bar{\Lambda}\bigr)=0,
\end{equation}
where the external part of $C_{4}$ has been written\footnote{We use
  notation which is slightly different than the remainder of the paper
  to match with the notation in~\cite{Giddings:2001yu} and to
  present~\eqref{eq:GKPequation} simply.}
\begin{equation}
  C_{4}=\alpha d x^{0}\wedge d x^{1}\wedge d x^{2}\wedge d x^{3},
\end{equation}
and
\begin{equation}
  \Lambda=\Phi^{+}G^{-}+\Phi^{-}G^{+}
\end{equation}
where $\Phi^{\pm}=h^{-1}\pm\alpha$ and $G^{\pm}=i
G_{3}\pm\tilde{\ast}_{6}G_{3}$.  Since in the KS background, $\tau$ is
constant and both $\Phi^{-}$ and $G^{-}$ vanish, the second term
in~\eqref{eq:GKPequation} is higher order in the perturbations and the
remainder of the equation implies $d\Lambda = 0$.  To leading
order in perturbations, this implies
\begin{equation}
  \label{eq:GKPeqperturbed}
  \Phi^{+}\delta G^{-}=-\delta\Phi^{-}G^{+}.
\end{equation}
Although the $\overline{\uD 3}$-branes do not directly couple to
$G^{\pm}$, they do directly source $\Phi^{-}$.  Since the KS geometry
has both $\Phi^{+}$ and $G^{+}$ non-vanishing, this direct coupling
implies that $G^{-}$ must be non-vanishing when an $\overline{\uD
  3}$-brane is added.  Furthermore, since $\delta\Phi^{-}$ will have
singular behavior at small $\tau$ while $\Phi^{+}$ and $G^{+}$ are
regular, $\delta G^{-}$ must have smaller powers of $\tau$ than
$G^{+}$.  For example, in the KS background $f-k\sim\tau$ while
$\Phi^{+}\sim\tau^{0}$.  Since $\delta\Phi^{-}\sim \tau^{-1}$, one
might expect $f_{p}-k_{p}\sim\tau^{0}$. Due to the presence of the
Hodge-$\ast$ which will introduce the squashing functions into this
analysis this argument alone is not conclusive and one must solve the
equations of motion as we did above. Nevertheless, it provides a
heuristic argument for why this singular behavior for $H_{3}^{2}$ and
$F_{3}^{2}$ is present in this solution.  The intuition that an
$\overline{\uD 3}$-brane should not result in such singular behavior
comes partially from the flat space case where $G^{+}=0$ and this
argument fails.  Similarly, it fails for the addition of $\uD
3$-branes to KS since $\uD 3$s source only $\Phi^{+}$ and not
$\Phi^{-}$.

The backreaction of $\overline{\uD 3}s$ was also addressed
in~\cite{Bena:2009xk}.  In~\cite{Bena:2009xk}, the existence of the
constant part of $k_{p}-f_{p}$ and the linear part of $F_{p}$, after
imposing that the more singular parts vanish, was deduced from a
slightly different logic.  The authors used the fact that a probe $\uD
3$-brane in the geometry will be attracted to $\overline{\uD
  3}$-branes at the tip\footnote{Of course, requiring the solution to
  exhibit non-SUSY behavior and that the warp factor behave as
  $\tau^{-1}$ will result in such a force.}.  Under the assumption
that the backreaction of the $\overline{\uD 3}$-branes could be
described as a linearized perturbation to the Klebanov-Strassler
geometry with at least some non-normalizable modes absent, it was
shown in~\cite{Bena:2009xk} that the existence of this force implies
such behavior.  It was then argued that this may imply that treating
the $\overline{\uD 3}$s-branes as a perturbation to the
Klebanov-Strassler background is not a valid procedure because the
$\overline{\uD 3}$-branes do not directly couple to $H_{3}$ and
$F_{3}$ and that therefore the resulting singular $H_{3}^{2}$ and
$F_{3}^{2}$ are unphysical.  The point of view that we adopt is that
although it is true that adding $\uD 3$-branes to KS or $\overline{\uD
  3}$-branes to flat space will not result in such behavior, in light
of~\eqref{eq:GKPequation} it is not surprising that such modes exist
when adding $\overline{\uD 3}$-branes to KS.  Therefore, unlike the
possibility discussed in~\cite{Bena:2009xk}, we do not impose that
$H_{3}^{2}$ and $F_{3}^{2}$ are non-singular.

%%%%%
\section{\label{sec:other-non-susy}Regular non-supersymmetric
  perturbations}
%%%%%

Here we present solutions which do not include a singular $O(1/\tau)$
behavior in the warp factor.  In this case, the warp factor is a power
series in $\tau$ consisting only of even powers.

%%%
\subsection{\label{sec:solut-with-regul}Unsquashed regular
  perturbations to KS}
%%%

The equations of motion~\eqref{eom} admit a solution that is regular
and unsquashed.  Taking the ansatz~\eqref{non-squashing perturbative
  ansatz}, we find for the dilaton and fluxes
\begin{subequations}
  \label{simplest regular solution}
\begin{align}
  \Phi_p=& {\cal P}, \notag\\
  F_p=&{\cal P} \left(- \frac{\tau^2}{6} - \frac{\tau^4}{180}-
    \frac{13\tau^6}{15120}\right) + {\cal T}\left(\frac{5 \tau^2}{48}
    +\frac{\tau^4}{288} +
    \frac{13 \tau^6}{24192}\right) ,\notag\\
  f_p=& {\cal P}\left( \frac{\tau^3}{12} - \frac{\tau^5}{240} +
    \frac{\tau^7}{1120}\right) +
  {\cal T}\left(- \frac{\tau^5}{192}+\frac{\tau^7}{2016}\right),\notag\\
  k_p=& {\cal P}\left(- \frac{\tau}{3} - \frac{7 \tau^3}{180}-
    \frac{11 \tau^5}{5040}\right) + {\cal T}\left(\frac{5 \tau}{12} +
    \frac{ \tau^3}{36} +
    \frac{5  \tau^5}{4032}\right),
\end{align}
while the warp factor is
\begin{equation}
  h_p = 2^{\frac{2}{3}} (g_s M \alpha')^2
  \varepsilon^{-{\frac{8}{3}}} \left[{\cal A} + {\cal T}\left(-
      \frac{5\tau^2}{6^{1/3} 24} + \frac{5 \tau^4}{ 6^{1/3}
        144}\right) \right].
\end{equation}
\end{subequations}
Again, these solutions are valid up to linear order in the parameters
$\mathcal{A}$, $\mathcal{P}$, and $\mathcal{T}$.  The solution
satisfies the dilaton equation~\eqref{eq:dilatoneom} up to
$O(\tau^4)$, the gravitational equations~\eqref{eq:einstein} up to
$O(\tau^2)$, the $H_3$ equations~\eqref{H3eom 1} and~\eqref{H3eom 2}
up to $O(\tau^3)$ and $O(\tau^5)$, and the $F_3$
equation~(\ref{F3eom}) up to $O(\tau^4)$.  It can be easily extended
to higher order in $\tau$ without introducing additional independent
parameters (i.e. higher order coefficients can be expressed in terms
of $\mathcal{P}$, $\mathcal{T}$, and $\mathcal{A}$).  The solution
related to the parameter ${\cal T}$ is the same solution appeared in
the exact solution~(\ref{Exact ISD solution with constant dilaton})
after expanding around $\tau=0$.
 
As was the case for the singular unsquashed perturbation given in
Sec.~\ref{sec:pert-solut-deform}, the fluxes in this solution respect
the ISD condition.  The solution has a non-vanishing
$\left(0,3\right)$-component
\begin{equation}
  G_3^{(0,3)} = \left(8 {\cal P} - 5{\cal T}\right)
  \left( \frac{1}{8\tau^2} - \frac{1}{24} + \frac{\tau^2}{120} - 
    \frac{\tau^4}{756} + \cdots \right)
  (z_i d\bar{z}_i)\wedge
  (\epsilon_{ijkl} z_i \bar{z}_j d\bar{z}_k \wedge d\bar{z}_l),
\end{equation}
while the $(3,0)$ and $\left(1,2\right)$-components vanish.  The
existence of the $\left(0,3\right)$ part implies that the gravitino
variation is non-vanishing for general choices of $\mathcal{P}$ and
$\mathcal{T}$ and thus supersymmetry is broken even though the flux is
ISD.  However, taking $8\mathcal{P}=5\mathcal{T}$ results in an
$\mathcal{N}=1$ supersymmetric solution.  This special case is a
generalization of KS, corresponding to a constant shift of the string
coupling and a canceling shift in $H_{3}$ such that $G_{3}$ is
unchanged.  Indeed in this case, $F_{p}=0$ while $f_{p}\propto f_{KS}$
and $k_{p}\propto k_{KS}$.

%%%
\subsection{\label{sec:squashedreg}Squashed regular perturbations to KS}
%%%

As was the case for the singular perturbations, it is possible to
obtain solutions that break the ISD condition by adopting the more
general squashed ansatz~\eqref{squashing perturbative ansatz}.  We
again find such a solution to~\eqref{eom} as a power series in $\tau$.
The dilaton profile is again non-trivial
\begin{subequations}
\label{regular general solution}
\begin{equation}
  \Phi_p =\varphi \left(-\frac{\tau ^2}{16}+\frac{\tau ^4}{96}-
    \frac{37 \tau ^6}{25200}+\frac{\tau ^8}{5250}\right).
\end{equation}
The metric squashing functions are
\begin{align}
  b_p =& {\cal D} \left(-3 \tau ^2+\frac{13 \tau ^4}{70}- \frac{517
      \tau ^6}{15750}\right)+{\cal M} \left(1-\frac{\tau ^2}{4}+
    \frac{\tau ^4}{42}-\frac{82 \tau ^6}{23625}\right)\notag\\
  & +{\cal Q} \left(-2 \tau ^2+\frac{\tau ^4}{30}-\frac{121 \tau
      ^6}{15750}\right) + \varphi \left(-\frac{\tau ^4}{560}+
    \frac{223 \tau ^6}{378000}\right), \notag\\
%\intertext{\newpage}
  q_p =& {\cal D} \left(\frac{3 \tau ^4}{28}-\frac{1847 \tau
      ^6}{126000}\right)+{\cal M} \left(1+\frac{\tau
      ^4}{112}-\frac{589 \tau
      ^6}{378000}\right)\notag\\
  &+{\cal Q} \left(\tau ^2-\frac{\tau ^4}{60}-\frac{101 \tau
      ^6}{126000}\right) +\varphi \left(-\frac{3 \tau
      ^4}{560}+\frac{1669 \tau ^6}{1512000}\right)
  , \notag\\
  s_p =& {\cal D}
  \left(\tau ^2-\frac{29 \tau ^4}{300}+\frac{13817 \tau ^6}{882000}\right)+
  {\cal M} \left(-\frac{3 \tau ^4}{400}+\frac{157 \tau ^6}{98000}\right)+
  {\cal Q} \left(\frac{\tau ^4}{100}+\frac{697 \tau ^6}{294000}\right)\notag\\
  &+\varphi \left(\frac{\tau ^4}{200}-\frac{1231 \tau ^6}{1176000}\right).
\end{align}
and the warp factor is
\begin{align}
  h_p =&(g_s M \alpha')^2 2^{\frac{2}{3}} \varepsilon^{-\frac{8}{3}}
  \left[{\cal D} \left(-\frac{13 \tau ^2}{5\ 6^{1/3}}+\frac{317\
        2^{2/3} \tau ^6}{7875\ 3^{1/3}}\right)\right.\notag \\
    & \hspace{8em}+{\cal M}\left.
    \left(-\frac{13 \tau ^2}
      {40\ 6^{1/3}}+\frac{1163 \tau ^6}{126000\ 6^{1/3}}\right)\right. \notag\\
  & \hspace{8em}\left. +{\cal Q} \left(-\frac{\tau ^2}{2\
        6^{1/3}}+\frac{13 \tau ^6}{1260\
        6^{1/3}}\right)\right. \notag \\
  & \hspace{8em} \left. \varphi\left(\frac{\tau ^2}{40\
        6^{1/3}}-\frac{23 \tau ^6}{18000\ 6^{1/3}}\right) \right].
\end{align}
The perturbed fluxes are
\begin{align}
  F_p =& {\cal D} \left(\frac{13 \tau ^2}{10}+\frac{17
      \tau^4}{75}-\frac{173 \tau ^6}{8400}+\frac{5921
      \tau ^8}{1323000}\right) \notag \\
  &+{\cal M} \left(\frac{139 \tau ^2}{240}+\frac{319 \tau
      ^4}{7200}-\frac{101 \tau ^6}{604800}+
    \frac{50087 \tau ^8}{127008000}\right)\notag\\
  &+{\cal Q} \left(\frac{\tau ^2}{4}+\frac{11 \tau ^4}{120}-\frac{27
      \tau ^6}{5600}+\frac{6457 \tau ^8}{6350400}\right)
  \notag\\
  &+\varphi\left(\left(-\frac{1}{80}-\frac{3^{1/3} a_0}{16\
        2^{2/3}}\right) \tau ^2+\left(\frac{19}{2400}-
      \frac{a_0}{160\  6^{2/3}}\right) \tau ^4\right.\notag\\
  &\quad\quad\left.+\left(-\frac{73}{67200}-\frac{13 a_0}{13440\
        6^{2/3}}\right) \tau ^6+\left(\frac{2059}{14112000}+\frac{a_0}
      {19200\ 6^{2/3}}\right) \tau ^8\right) ,  \notag\\
%\intertext{\newpage}
  f_p = &{\cal D} \left(-\frac{3^{1/3} a_0 \tau ^3}{2 2\
      ^{2/3}}+\left(\frac{11}{100}-\frac{3^{1/3} a_0}{40 \
        2^{2/3}}\right) \tau ^5+\left(-\frac{29}{800}-\frac{a_0}{560\
        6^{2/3}}\right)
    \tau ^7\right. \notag\\
  & \left. \quad\quad +\left(\frac{164063}{21168000}-\frac{a_0}{40320\
        6^{2/3}}\right) \tau
    ^9\right)\notag\\
  &+{\cal M} \left(\left(-\frac{1}{8}+\frac{a_0}{8\ 6^{2/3}}\right)
    \tau ^3+\left(-\frac{13}{1600}+\frac{a_0}{160\ 6^{2/3}}\right)
    \tau
    ^5\right.\notag\\
  &\quad\quad\left.+\left(-\frac{57}{22400}+\frac{a_0}{6720\
        6^{2/3}}\right) \tau
    ^7+\left(\frac{314123}{508032000}+\frac{a_0}{483840\
        6^{2/3}}\right) \tau
    ^9\right)\notag\\
  &+{\cal Q} \left(-\frac{\tau
      ^5}{80}-\frac{\tau ^7}{700}+\frac{22003 \tau ^9}{42336000}\right) \notag\\
  &+\varphi\left(\frac{a_0 \tau ^3}{32\
      6^{2/3}}+\left(-\frac{17}{4800}-\frac{a_0}{128\ 6^{2/3}}\right)
    \tau ^5+\left(\frac{359}{201600}+\frac{5
        a_0}{5376\  6^{2/3}}\right) \tau ^7\right. \notag\\
  &\quad\quad\left.+\left(-\frac{227047}{508032000}-\frac{233 a_0}
      {1935360\  6^{2/3}}\right) \tau ^9\right) ,\notag\\
\intertext{\newpage}
  k_p =&{\cal D} \left(\left(\frac{26}{5}+3\ 6^{1/3} a_0\right) \tau
    +\left(\frac{47}{150}+\frac{3^{1/3} a_0}{2\ 2^{2/3}}\right) \tau
    ^3+\left(\frac{11}{1050}+\frac{3^{1/3}
        a_0}{40\  2^{2/3}}\right) \tau ^5 \right. \notag\\
  & \quad\quad \left. +\left(\frac{289}{1764000}+\frac{a_0}
      {560\  6^{2/3}}\right) \tau ^7\right)\notag\\
  &+{\cal M} \left(\left(\frac{149}{60}-\frac{3^{1/3} a_0}{2\
        2^{2/3}}\right) \tau +\left(\frac{209}{900}-\frac{a_0}{8\
        6^{2/3}}\right) \tau ^3\right. \notag\\
  &\quad\quad \left. +\left(\frac{979}{100800}-\frac{a_0}{160\
        6^{2/3}}\right) \tau
    ^5+\left(\frac{311}{1411200}-\frac{a_0}{6720\ 6^{2/3}}\right)
    \tau ^7\right)\notag\\
  &+{\cal Q} \left(\tau +\frac{2 \tau ^3}{5}+\frac{109 \tau
      ^5}{8400}-\frac{19 \tau ^7}{496125}
  \right)\notag\\
  &+\varphi\left(\left(-\frac{1}{20}+\frac{3^{1/3} a_0}{8\
        2^{2/3}}\right) \tau +\left(-\frac{1}{50}+
      \frac{3^{1/3} a_0}{160\ 2^{2/3}}\right) \tau ^3\right. \notag\\
  &\quad\quad \left. +\left(-\frac{1}{4800}+\frac{3^{1/3} a_0}{4480\
        2^{2/3}}\right) \tau ^5+\left(-\frac{1433}{63504000}+
      \frac{a_0}{44800\ 6^{2/3}}\right) \tau ^7\right),
\end{align}
\end{subequations}
where we have again omitted terms presented in
Sec.~\ref{sec:solut-with-regul}.  This solutions is valid to linear
order in the parameters $\varphi$, $\mathcal{M}$, $\mathcal{Q}$, and
$\mathcal{D}$ which characterize the perturbation and again one could
extend this to higher orders in $\tau$.

The resulting $G_{3}$ is no longer purely ISD since
\begin{equation}
  H_3^2 - e^{2\Phi} F_3^2 = \frac{6^{1/3}9 \varphi}
  {a_0^{1/2} g_s^3 M \alpha'} +
  \frac{3(-1+6^{1/3} 4 a_0) \varphi \tau^2}{2 a_0^{3/2} g_s^3 M \alpha'}
  +\cdots\neq0.
\end{equation}
This can also be checked more directly using~\eqref{ISD & IASD
  condition}.  Although only the parameter $\varphi$ appears in the
potential for the dilaton, making any of these independent parameters
non-zero leads to a non-ISD flux\footnote{The vanishing of the
  potential~\eqref{eq:dilatonpotential} merely implies
  $\mathrm{Re}\left(G^{+}_{mnp}G^{-mnp}\right)=0$, not that
  $G^{+}=0$.}.
 
%%%%%
\section{\label{sec:comments-non-susy}Non-SUSY solutions in the KT
  region}
%%%%%

It is also possible to find non-SUSY perturbations to the KT solution.
We again will find that non-ISD fluxes can be found only if the
conifold is squashed.  As before, we consider solutions that are
linear in the perturbations, though since we are working at
large $\tau$, we do not perform a power series expansion around
$\tau=0$.

%%%
\subsection{\label{sec:ktsol}Klebanov-Tseytlin solution}
%%%

The ansatz~\eqref{geometry ansatz},~\eqref{gauge ansatz} also includes
the KT solution~\cite{Klebanov:2000nc}.  This solution corresponds to
adding $N$ $\uD 3$-branes and $M$ fractional $\uD 3$-branes (i.e. $M$
$\uD 5$-branes wrapping a collapsing 2-cycle) to the undeformed
conifold singularity and is valid at large distances from the conifold
point.  It is recovered by
\begin{align}
  f_{KT}(r) =& k_{KT}(r) = \frac{3}{2}\log\frac{r}{r_0},\quad
  F_{KT}(r) = \frac{1}{2},\notag \\
  \ell_{KT}(r)=& f_{KT}(1-F_{KT}) + k_{KT} F_{KT}+
  \frac{\pi N}{g_s M^2},\notag\\
  p_{KT}(r)=& 1,\quad b_{KT}(r) = \frac{r^2}{9},\quad
  q_{KT}(r) = s_{KT}(r) = \frac{r^2}{6},\notag\\
  \Phi_{KT}(r) =& \log g_s,\quad h_{KT}(r) = \frac{27
    \pi}{4r^4}\left(g_s N \alpha'^2 +\frac{3}{8\pi} (g_s M \alpha' )^2 + \frac{3}{2\pi} (g_s M \alpha'
    )^2 \log \frac{r}{r_0} \right),
  \label{KT solution}
\end{align}
where $r^{2}=2^{5/3}3\ve^{4/3}e^{2\tau/3}$.  In contrast to the KS
solution, $\ell$ is chosen to satisfy
\begin{equation}
  {\cal F}_5 = 27 \pi \alpha'^2 N \mathrm{vol}_{T^{1,1}} + B_2 \wedge F_3,
\end{equation}
where $\mathrm{vol}_{T^{1,1}}$ is the volume form of the angular
space.  This reflects the fact that the effective $\uD 3$ charge
receives contributions from both the $3$-form fluxes and the $N$
regular $\uD 3$-branes which provide a localized source for the
charge.

%%%
\subsection{Unsquashed perturbations to KT}
%%%

In analogy with the analyses of unsquashed perturbations of KS in
Sections~\ref{sec:pert-solut-deform} and~\ref{sec:solut-with-regul},
we first consider perturbations for which the unwarped 6D space is
still the unsquashed conifold and the flux is ISD.  We take the
ansatz
\begin{align}
  \Phi =& \log g_s + \Phi_{p}(r),\quad h = h_{KT} + h_p(r),\quad
  \ell = f(1-F) + k F+ \frac{\pi N}{g_s M^2},\notag\\
  f=& f_{KT} + f_{p}(r),\quad k=k_{KT} + k_{p}(r),\quad
  F=F_{KT} + F_{p}(r),\notag\\
  p=&p_{KT},\quad b=b_{KT} ,\quad q=q_{KT},\quad s=s_{KT}.
\end{align}
Solving the ISD condition~\eqref{ISD & IASD condition} and the first
order equation~(\ref{eq:warpfactorsusy}) yields
\begin{align}
  \Phi_p = & {\cal P} + \frac{\phi}{r^4},\notag\\
  F_p =& \frac{{\cal G}}{r^3} ,\notag\\
\intertext{\newpage}
  f_p =& {\cal C} + \frac{{\cal G}}{r^3} - \frac{3 \phi}{8 r^4} +
  \frac{3 {\cal P}}{2}
  \log \frac{r}{r_0},\notag\\
  k_p =& {\cal C} - \frac{{\cal G}}{r^3} - \frac{3 \phi}{8 r^4}+
  \frac{3 {\cal P}}{2}
  \log \frac{r}{r_0},\notag\\
  h_p =& (g_s M \alpha')^2 \left[{\cal A} + \frac{27 {\cal C}}{r^4} +
    \frac{81 {\cal P}}{8 r^4}\left(\frac{1}{4} + \log \frac{r}{r_0}
    \right)- \frac{81 \phi}{64 r^8}\right],
  \label{KT perturbation with ISD}
\end{align}
where we have retained only solutions that are regular as
$r\to\infty$.  The solution is valid to linear order in the parameters
${\cal P}$, $\phi$, ${\cal C}$, ${\cal G}$, and ${\cal A}$ which
characterize the perturbation.  Note that some terms in the
perturbation are sub-dominant to the corrections to the KT geometry
coming from the full KS solution; however even if these corrections
are included, the perturbations are not corrected until even higher
order in $1/r$.
  
The parameters ${\cal P}$ and $\phi$ are essentially the same
parameters that appear in the perturbations to KS in
Sections~\ref{sec:solut-with-regul} and~\ref{sec:pert-solut-deform}
respectively.  That is, ${\cal P}$ is a constant shift of the string
coupling and the part including $\phi$ is a solution to Laplace's
equation $\nabla^{2}\Phi=0$.  The parameters $\mathcal{G}$ and
$\mathcal{U}$ are related to those appearing in~\eqref{singular
  simplest solution} and~\eqref{Exact ISD solution with constant
  dilaton} as $\mathcal{G}=2\ve^{2}\mathcal{U}$ and
$\mathcal{C}=\frac{\mathcal{H}}{6}-\frac{8\mathcal{U}}{3}$ (the
remaining parameter $\mathcal{T}$ appearing in~\eqref{Exact ISD
  solution with constant dilaton} is not regular as $r\to\infty$).

The parameter $\phi$ is also the same parameter appearing
in~\cite{DeWolfe:2008zy}.  By calculating the Hawking-Horowitz
mass~\cite{Hawking:1995fd} (the generalization of ADM mass), which is valid at large radius, the
authors of~\cite{DeWolfe:2008zy} concluded that the relevant behavior
of the perturbation to the warp factor due to the $\uD
3$-$\overline{\uD 3}$ pairs should include a term behaving as
$r^{-8}\log r$.  However no such a term appears in~\eqref{KT
  perturbation with ISD}.  Moreover there is no squashing and the flux
remains ISD.  Therefore, even though SUSY is broken in this solution,
it does not correspond to the presence of $\overline{\uD 3}$-branes.

%%%
\subsection{Squashed perturbations to KT}
%%%

A perturbation of KT which is no longer ISD was found
in~\cite{DeWolfe:2008zy}.  Based on a similar analysis of
$\mathrm{AdS}_{5}\times {\mathbb S}^{5}$, the authors
of~\cite{DeWolfe:2008zy} assume the perturbations due to the
$\overline{\uD3}$-branes behave as $O(r^{-4}, r^{-4}\log r)$ relative
to the original KT solution and took an ansatz which squashes each of
the $SU(2)$-isometry directions in the same way.  However, it is
interesting to relax this condition and take the more general
ansatz~(\ref{geometry ansatz}),~(\ref{gauge ansatz}) with
\begin{align}
  \Phi =& \log g_s + \Phi_{p}(r),\quad h = h_{KT} + h_p(r),\quad
  \ell = f(1-F) + k F+ \frac{\pi N}{g_s M^2},\notag \\
  f=& f_{KT} + f_{p}(r),\quad k=k_{KT} + k_{p}(r),\quad
  F=F_{KT} + F_{p}(r),\notag\\
  b=&b_{KT}(1 + b_p(r)),\quad q = q_{KT}(1+ q_p(r)), \quad s= s_{KT}(1
  + s_p(r)),\quad p=p_{KT}.
  \label{squashing perturbative ansatz 2}
\end{align}
Such an ansatz in general squashes the spheres in different ways.
Assuming perturbations that behave as $O(1, \log r, r^{-4}, r^{-4}\log
r)$ relative to the KT solution\footnote{As was the case in the
  previous section, including the finite deformation corrections to
  KT will not change the form of the perturbations.}, the equations of
motion~\eqref{eom} admit a solution
\begin{align}
  \label{eq:squashedktpert}
  \Phi_p =& - \frac{3 {\cal S} \log(r/r_0)}{r^4},\notag\\
  b_p =& {\cal J} + \frac{{\cal S}}{r^4},
  \quad q_p = s_p = {\cal J}\notag\\
  k_p =& f_p = \frac{{\cal S}}{r^4} \left(\frac{33}{32} + \frac{3 N
      \pi}{4 g_s M^2} + \frac{9}{4} \log\frac{r}{r_0}\right),
  \quad F_p =0,\notag\\
  h_p =& - \frac{27\pi {\cal J}}{ 2r^4} \left(g_s N \alpha'^2 +
    \frac{3}{8\pi}(g_s M \alpha')^2 + \frac{3}{2\pi}(g_s M \alpha')^2
    \log \frac{r}{r_0}\right),\notag\\
  &+ \frac{{\cal S}}{r^8}\left(\frac{27\pi}{32}g_s N \alpha'^2 +
    \frac{1053}{256}(g_s M \alpha')^2 + \frac{81}{16}(g_s M \alpha')^2
    \log \frac{r}{r_0}\right),
\end{align}
where ${\cal J}$ and $\mathcal{S}$ parameterize the perturbation and
we omit the parameters which have appeared in the previous
subsection. The parameter $\mathcal{S}$ is the same parameter
appearing in~\cite{DeWolfe:2008zy} and breaks the ISD condition and
thus breaks SUSY.  It was shown in~\cite{DeWolfe:2008zy} that
$\mathcal{S}$ contributes a finite amount to the ADM mass as one would
expect from the addition of $\uD 3$s or $\overline{\uD 3}$s but since
it does not contribute to the net charge, $\mathcal{S}$ characterizes
the influence of $\uD 3$-$\overline{\uD 3}$ pairs.

Similarly, while turning on the parameter $\mathcal{J}$ preserves the
ISD condition (\ref{ISD & IASD condition}) and the first derivative
equation for warp factor (\ref{eq:warpfactorsusy}) and does not
introduce a $\left(0,3\right)$-component to $G_{3}$, it causes the
unwarped 6D space to no longer be Ricci flat (and therefore no longer
Calabi-Yau) so that there is no spinor covariantly constant with
respect to the unwarped metric, implying that supersymmetry is
broken. In order for the flux part of the Killing spinor equations to
vanish, any Killing spinor of the perturbed geometry would have to
satisfy the same chirality conditions as the Killing spinor of the
unperturbed geometry (i.e.  $\Gamma_{z}\epsilon=0$ where $z$ is any
holomorphic coordinate of KT).  Therefore, while the flux part of the
SUSY variation of the gravitino vanishes, the spin connection part
does not and SUSY is broken.  A priori, one might expect a
cancellation between the flux and spin connection parts might be
possible for a different choice of chirality, but one can show that
this cannot occur (see e.g.~\cite{Grana:2005jc} and references
therein).

Although a non-vanishing $\mathcal{J}$ breaks supersymmetry, it does
not describe the presence of $\overline{\uD 3}$-branes.  Note that
while taking $\mathcal{J}\neq 0$ does not add any charge to the
background, it still might describe the presence of $\uD
3$-$\overline{\uD 3}$-brane pairs.  However, such a configuration
would still provide a localized source of tension.  The constant shift
$\mathcal{J}$ in the squashing cannot be a result of a localized
tension since such a source should cause a functional form that is
singular as $r\to 0$.  Similarly, the perturbed warp factor is not a
result of additional localized sources of tension, but results as a
solution of~\eqref{eq:warpfactorsusy} with the perturbed squashing
functions.  Thus, the large radius backreaction of $\uD
3$-$\overline{\uD 3}$-pairs is found by setting $\mathcal{J}=0$,
reproducing the result of~\cite{DeWolfe:2008zy}\footnote{We thank
  S.~Kachru and M.~Mulligan for some useful comments related to this
  discussion.}.
 
%%%%%
\section{\label{sec:gravitino-mass-}Gravitino mass}
%%%%%

In this section we calculate the effective 4D gravitino mass that
results from dimensional reduction of the SUSY breaking solution.  The
gravitino can potentially obtain a mass from interactions with the
$5$-form flux $F_{5}$ and the $3$-form flux $G_{3}$.  This problem was
addressed previously in~\cite{DeWolfe:2002nn}, though in their
analysis they considered a background for which the warp factor
satisfied the condition~\eqref{eq:warpfactorsusy}.  However, even when
this condition is not satisfied, their method can still be applied and
we follow it closely here.  Note that although we are interested in
the specific case of the warped deformed conifold, this discussion
applies to any perturbation of a warped Calabi-Yau.
 
Since we work in the Einstein frame, we relate the Einstein frame
spinors to those in the string frame
\begin{align}
  \Psi^E_M =& g_s^{\frac{1}{8}}e^{-\frac{\phi}{8}}\Psi_M^s -
  \frac{i}{4} g_s^{\frac{1}{8}}e^{-\frac{\phi}{8}}
  \Gamma_M\lambda^{s*},
  &\lambda^E=&g_s^{-\frac{1}{8}}e^{\frac{\phi}{8}}\lambda^s, \notag \\
  \Gamma_M^E =& g_s^{\frac{1}{4}} e^{-\frac{\phi}{4}}\Gamma_M^s,
  &\epsilon^E =& g_s^{\frac{1}{8}} e^{-\frac{\phi}{8}}\epsilon^s.
\end{align}
Up through bilinear terms, the action for the type IIB Einstein frame
gravitino is
\begin{align}
  S_{f}=& \frac{1}{\kappa^2}
  \int d^{10} x \sqrt{-g}\bigl(\mathcal{L}_{1}+\mathcal{L}_{2}\bigr), \notag \\
  \mathcal{L}_{1}= &i \bar{\Psi}_M \Gamma^{MNS} \left(D_N \Psi_S +
    \frac{i}{4} e^{\Phi} \partial_N C \Psi_S +
    i\frac{g_s}{192}\Gamma^{R_1R_2R_3R_4}
    F_{S R_1R_2R_3R_4} \Psi_N \right), \notag\\
  \mathcal{L}_{2}=& -i \frac{g_{s}^{1/2} e^{\Phi / 2}}{192}
  \bar{\Psi}_M \Gamma^{MNS} \left( \Gamma_{S}^{\
      R_1R_2R_3}G_{R_1R_2R_3} - 9 \Gamma^{R_1R_2} G_{S R_1R_2}\right)
  \Psi_N^* + {\rm h.c.},
  \label{gravitino action}
\end{align}
where $D_{M}$ is the covariant derivative which, when acting on
$\Psi_{\mu}$, is given by
\begin{align}
  D_{[\mu} \Psi_{\nu]}=&\hat{D}_{[\mu} \Psi_{\nu]} - \frac{1}{8}
  \Gamma_{[\mu} \Slash{\partial} \log h \, \Psi_{\nu]},\notag \\
  D_{[m} \Psi_{\nu]}=&\hat{D}_{[m} \Psi_{\nu]} + \frac{1}{8}
  \Gamma_{[m} \Slash{\partial} \log h \, \Psi_{\nu]} -
  \frac{1}{8} \partial_{[m} \log h \, \Psi_{\nu]},
\end{align}
where $\hat{D}_{\mu}$ and $\hat{D}_{m}$ are the covariant derivatives
built from the unwarped metrics $\hat{g}_{\mu\nu}$ and $\hat{g}_{mn}$.

The $\Psi_{\mu}$ part of the 10D gravitino is decomposed as a product
of a 4D gravitino $\psi_{\mu}$ and a 6D spinor $\chi$ that is
covariantly constant with respect to the unwarped metric
\begin{align}
  \Psi_\mu\left(x^{\mu},x^{m}\right)= \psi_\mu\left(x^{\mu}\right)
  \otimes h^{-\frac{1}{8}} \chi\left(x^{m}\right),
\end{align}
where $\chi$ is normalized such that $\chi^\dagger \chi = 1$.  The
$h^{-1/8}$ factor of the warp factor comes from requiring that the
spinor is covariantly constant with respect to the warped metric,
$\hat{D}_{m}\Psi_\mu =0$~\cite{Grana:2001xn}.

The 4D kinetic term following from~\eqref{gravitino action} can be
evaluated by dimensional reduction
\begin{equation}
  \frac{1}{\kappa^2}\int d^{10} x \sqrt{-g}\, i \bar{\Psi}_\mu
  \Gamma^{\mu\nu\rho} \hat{D}_\nu \Psi_\rho
  = \frac{1}{\kappa_4^2} \int d^4 x \sqrt{-\hat{g}_4} \, i \bar{\psi}_\mu
  \hat{\Gamma}^{\mu\nu\rho} \hat{D}_\nu \psi_\rho,
\end{equation}
where on the right hand the indices are contracted with the unwarped
metric $\hat{g}_{\mu\nu}$ and where the 4D gravitational constant and
the warped volume are
\begin{equation}
  \frac{1}{\kappa_4^2} \equiv\frac{1}{\kappa^2} \vol_6^w,\qquad
  \vol_6^w \equiv \int d^6 y \sqrt{\hat{g}_6}\, h.
\end{equation}

If the supersymmetry condition on the warp
factor~\eqref{eq:warpfactorsusy} is satisfied, then the coupling to
$F_5$ is canceled by the spin connection.  However in general this
interaction term could a priori contribute to the gravitino mass and
we have
\begin{multline}
  \frac{1}{\kappa^2}\int d^{10} x \sqrt{-g} \,i \bar{\Psi}_M
  \Gamma^{MNR}\left( - \frac{1}{4}\omega_R^{AB} \hat{\Gamma}_{AB} + i
    \frac{g_s}{16}
    \Slash{F}_5 \Gamma_R \right) \Psi_N \\
  \ni - \frac{1}{\kappa^2} \int d^4 x \sqrt{-\hat{g}_4} \bar{\psi}_\mu
  \hat{\Gamma}^{\mu\nu}\psi_\nu \int d^6 y \sqrt{\hat{g}_6} \frac{i}{8
    h^{1/2}} \left(\frac{h'}{\sqrt{p}} + \frac{(g_s M \alpha'^2)}{4}
    \frac{\ell}{\sqrt{b} q s}\right) \chi^\dagger \hat{\Gamma}_\tau
  \chi,
\end{multline}
where $\omega_{M}^{AB}$ is the spin connection with letters from the
beginning of the alphabet denoting tangent space indices and where on
the right hand side, terms involving the unwarped spin-connections
have been omitted and indices are again contracted with the unwarped
metric.  The gravitino mass resulting from the $5$-form flux is then
\begin{align}
  \frac{i}{8 V_6^w} \int d^6 y \sqrt{\hat{g}_6} h^{-1/2}
  \left(\frac{h'}{\sqrt{p}} + \frac{(g_s M \alpha'^2)}{4}
    \frac{\ell}{\sqrt{b} q s}\right) \chi^\dagger \hat{\Gamma}_\tau
  \chi.
\end{align}
However, this term vanishes as a result of the 6D chirality of $\chi$
and thus $F_{5}$ does not contribute to the gravitino mass.

The essential contribution to the gravitino mass comes from the
$3$-form flux.  Dimensional reduction gives
\begin{equation}
  \frac{1}{\kappa^2} \int d^4 x \sqrt{-\hat{g}_4} \,
  \bar{\psi}_\mu \hat{\Gamma}^{\mu\nu} \psi_\nu^*
  \int d^6 y \sqrt{\hat{g}_6} \left( \frac{i \sqrt{g_s} e^{\Phi/2}}{64}
    \chi^\dagger \hat{\Gamma}^{m n p} \chi^* G_{m n p}+ {\rm h.c.}\right).
\end{equation}
Since $\hat{\Gamma}^{\bar{\imath}} \chi = 0$, we can write
\begin{equation}
  \chi^\dagger \hat{\Gamma}^{m n p} \chi^*=
  \chi^\dagger \hat{\Gamma}^{\bar{\imath}\bar{\jmath}\bar{k}} \chi^* = 
  \Omega^{\bar{\imath}\bar{\jmath}\bar{k}},
\end{equation}
where $\Omega$ is the holomorphic $3$-form of the underlying
Calabi-Yau whose explicit form for the deformed conifold is given
in~\eqref{holomorphic 3-form}.  Thus only the
$\left(0,3\right)$-component of $G_{3}$ contributes to the gravitino
mass\footnote{We are treating the background as a non-SUSY
  perturbation to a warped Calabi-Yau.  More generally, when the
  Calabi-Yau is squashed there will be additional potential
  contributions from terms such as $g^{ij}g^{kl} g^{\bar{m}n}
  G_{ik\bar{m}} \Omega_{jln}$, but these are higher order in
  perturbations since the unperturbed metric has
  $g_{ij}=g_{\bar{\imath}\bar{\jmath}}=0$.}.  This has been shown
previously~\cite{DeWolfe:2002nn}, but here we argued that it holds
even when~\eqref{eq:warpfactorsusy} is not satisfied.  The 4D
gravitino mass resulting from the $3$-form flux is then
\begin{align}
  m_{3/2} =& \frac{3\sqrt{g_s}}{\,i {\cal V}_6^w} \int e^{\Phi/2} \, \Omega
  \wedge G_3,
  \label{gravitino mass GVW superpotential form}
\end{align}
which is quite similar to what follows from the Gukov-Vafa-Witten
superpotential~\cite{Gukov:1999ya}.  With the explicit formula for the
of K\"ahler potential\footnote{Here we continue to
  follow~\cite{DeWolfe:2002nn}, but in the presence of strong warping,
  the K\"ahler potential should be modified from the expression used
  there~\cite{Giddings:2005ff,Shiu:2008ry,Douglas:2008jx,Frey:2008xw,%
    Marchesano:2008rg,Martucci:2009sf,Chen:2009zi}.}
and restoring the K\"ahler modulus $\rho$, we can write the gravitino
mass as~\cite{DeWolfe:2002nn}
\begin{align}
  m_{3/2} \propto \kappa_4^2 e^{\frac{\cal K}{2}} W_{GVW},
\end{align}
where $W_{GVW}$ is the GVW superpotential and $\mathcal{K}$ is the
K\"ahler potential.

If we apply these expressions for the gaugino mass to~\eqref{singular
  general solution}, we find
\begin{equation}
  \label{eq:gravitinomass1}
  m_{3/2}\sim \kappa_4^2 \frac{\left(\mathcal{S}+10\mathcal{T}\right)\ve^{2/3}}
  {a_{0}\left(g_{s}M\alpha'\right)\tau_{\mathrm{min}}},
\end{equation}
In evaluating this, we have assumed that most of the contribution to
the gravitino mass should come from small $\tau$, close to where the
source of SUSY breaking is located, and cut the integral at some lower
bound $\tau_{\mathrm{min}}$.  The lower bound must be introduced
because for sufficiently small $\tau$, the supergravity approximation
breaks down.  For the singular solutions of
Sec.~\ref{sec:non-susy-deformation} where the warp factor behaves at
small $\tau$ as $O(1/\tau)$, the Ricci scalars of these backgrounds
behave as $R \sim {\cal S} / (g_s M \alpha' \tau)$ where ${\cal S}$
stands for any of the parameters characterizing the perturbation
(which we expect to be all of the same order for a given solution).
Thus, the solutions are valid for $\tau$ satisfying $1/(g_s M \alpha')
\ll \tau < 1$.  If we na\"ively take $\tau_{\mathrm{min}}$ to be this
lower bound then
\begin{equation}
  m_{3/2}\sim \kappa_{4}^{2}\mathcal{S}\ve^{2/3}.
\end{equation}
This is a finite value even if $g_s M$ is large.  A more precise
calculation of the gravitino mass would require extending the integral
to smaller $\tau$ where the stringy corrections to the geometry become
important.
 
We also found solutions which behaves regularly at $\tau=0$.  The
result of the calculation for the solution in Section~\ref{sec:squashedreg}
is
\begin{align}
  \label{eq:gravitinomass2}
  m_{3/2}\sim\kappa_{4}^{2}\frac{\ve^{2/3}}{g_{s}M\alpha'}
  \bigl[&\bigl(-318+20\, 6^{1/3}a_{0}\bigr)\mathcal{M}
  -120\mathcal{Q} \notag \\
  &-\bigl(624+240\, 6^{1/3}a_{0}\bigr)\mathcal{D} +\bigl(6+ 5\,
  6^{1/3}a_{0}\bigr)\varphi\bigr]
\end{align}
This is a finite value, but since $\mathcal{S}$ is taken to be
perturbatively small, and $g_{s}M$ is large, the mass of the gravitino
is highly suppressed.

The solutions~\eqref{singular simplest solution}
and~\eqref{sec:solut-with-regul} yield values for the gravitino mass
that are similar to~\eqref{eq:gravitinomass1}
and~\eqref{eq:gravitinomass2} respectively.

%%%%%
\section{\label{sec:discussions}Discussion}
%%%%%

In this paper, we analyze several solutions to type IIB supergravity,
corresponding to non-supersymmetric perturbations to the warped
deformed conifold.  Of particular interest are the solutions presented
in Sec.~\ref{sec:squashedsingularks} which capture some key properties
of a solution describing the backreaction of $\overline{\uD 3}$-branes
smeared over the finite $\mathbb{S}^{3}$ at $\tau=0$.  In particular,
we discussed the {\it necessary} boundary conditions in the IR for the
solution to describe a localized $\overline{D3}$ source and how these
IR boundary conditions lead to the constant component of $H_{3}$ that
was discussed in~\cite{Bena:2009xk}. These solutions are thus related
to a small $\tau$ expansion of a background whose large radius
behavior was found in~\cite{DeWolfe:2008zy} and is dual to a
metastable SUSY breaking state.

For all of the above solutions, we have assumed the validity of a
linearized approximation.  For a small number $\overline{\uD
  3}$-branes, it is natural to expect that the linearized
approximation is valid at least at large distances where the
background flux largely dominates the effects of the $\overline{\uD
  3}$, though an extrapolation to larger radii would be necessary to
confirm this. For small distances, one can ensure that the linearized
approximation is good for $\tau$ above some particular value
determined by the parameters of the solution.  The linearized
approximation requires, for example that $F_{p}\ll F_{KS}$.  Using the
perturbations of Sec.~\ref{sec:squashedsingularks} and taking
$\mathcal{S}\sim\mathcal{B}\sim\mathcal{Y}$, this gives the condition
$\tau \gg {\cal S}^{1/3}$ where ${\cal S} \sim \kappa_{10}^2 T_3
(N_{\uD 3}+N_{\overline{\uD 3}})/(V_2 g_s^2 M^2 \alpha'^2)$.  Similar
or less restrictive conditions follow by considering the other
functions in the perturbation.  As discussed above, a similar
constraint is imposed by demanding that the
curvature~\eqref{eq:curvature} is small in string units\footnote{The
  additional requirement that $h_{p}\lesssim h_{p}^{2}$ can be
  satisfied if $\ve$ is not too small}.  Note that for large $M$,
$\tau$ is allowed to be quite small.  For the other solutions
presented above for which there is not always an obvious boundary
condition to impose, the validity of the linearized approximation is
more difficult to check.

There are several remaining open lines of research.  A particularly
important remaining open problem is to find a solution that
interpolates between the small and large radius regions\footnote{As
  mentioned in the introduction, some progress was made in this
  direction after this paper was completed~\cite{Bena:2009xk}.}.  Such
a solution would be important for many reasons.  For example, all of
the above solutions should admit a dual description as either
deformations of the KS gauge theory or states in the (possibly
deformed) KS gauge theory.  Although for some of the solutions the
field theory interpretation has been studied (for example, the dual of
the $\overline{\uD 3}$ solution was considered
in~\cite{DeWolfe:2008zy}), analysis of the remaining solutions would
clearly require extrapolating them to the UV.  Additionally, the
boundary conditions discussed in Section~\ref{sec:boundary-condition}
do not seem to be sufficient to fix all of the integration constants.
Having a solution that is valid at all distances would allow for a
calculation of quantities such as the Hawking-Horowitz mass or the
asymptotic charge which could provide other conditions to fix the
integration constants.  Finally, an interpolating solution would allow
for a more precise calculation of the flux-induced gravitino mass and
similar quantities.  Unfortunately, even the linearized equations of
motion are likely too complex to solve analytically in which case the
solution could only be presented numerically or formally in terms of
integrals, an analysis that we leave for future work.

The solutions could be improved in other ways.  For example, the
solutions presented in Sections~\ref{sec:pert-solut-deform}
and~\ref{sec:squashedsingularks} exhibit curvature singularities as
$\tau\to0$ and it is an interesting, though difficult, problem to
understand the stringy modifications of those backgrounds.  More
modestly, it would be interesting to relax the assumption that the
solutions retain the same isometry as the KS solution by, for example,
not smearing the $\overline{\uD 3}$-branes over the ${\mathbb S}^3$
\footnote{The localization of three-branes was considered in \cite{Klebanov:2007us} in different context.}.
One can also consider similar perturbations to the baryonic branch
solution \cite{Butti:2004pk,Maldacena:2009mw}.

Along similar lines, the solution~\eqref{singular general solution},
which, for some choice of parameters, would describe the effect of
$\overline{\uD 3}$-branes on the near tip geometry of KS, has been
argued to be a metastable background~\cite{Kachru:2002gs}.  However,
it would be interesting to use the explicit solution to analyze
fluctuations about this geometry to confirm the perturbative
stability, though this would require moving beyond the linearized
approximation.

Our solutions have potential applications to model building in warped
compactifications.  For example, the addition of $\overline{\uD
  3}$-branes into the warped deformed conifold was an important step
in the construction of stabilized de Sitter
%vacua in
vacua~\cite{Kachru:2003aw} and
%to
in the
%model
modeling of inflation
(see~\cite{McAllister:2007bg,HenryTye:2006uv,Cline:2006hu,Kallosh:2007ig,Burgess:2007pz}
and references therein).  It would be interesting to understand the
impact of the backreaction of the $\overline{\uD 3}$-branes on these
scenarios.  The construction in~\cite{Kachru:2003aw} further inspired
the scenario of mirage mediation~\cite{Choi:2005uz} and one might use
the solutions given here to provide a more string theoretical
understanding of this scheme.

A related though conceptually distinct application is in the context
of gauge-gravity duality.  The large radius
solution~\cite{DeWolfe:2008zy} was used in~\cite{Benini:2009ff} as a
holographic dual of a metastable SUSY breaking state.  The large
amount of isometry in this large radius region was found to suppress
gaugino masses in their construction.  However, the small radius
solution presented in Sec.~\ref{sec:squashedsingularks}, has reduced
isometry, and should result in more significant contributions.
Details of the application to holographic gauge mediation will be
discussed in a companion paper~\cite{McGuirk:2009am}.

\section*{Acknowledgments}
We have benefited from discussions with A. Dymarsky, A.~Ishibashi,
S.~Kachru, I. R. Klebanov, F.~Marchesano, L.~McAllister, M.~Mulligan,
P.~Ouyang, Y.~Tachikawa, and H. L. Verlinde.  PM and GS also thank
T. Liu for collaboration and preliminary discussions on this and
related topics. We would like to thank the Institute for Advanced
Study and the Hong Kong Institute for Advanced Study, Hong Kong
University of Science and Technology for hospitality and support. PM
and GS also thank the Standford Institute for Theoretical Physics and
SLAC for hospitality while some preliminary discussions were held.  YS
appreciates the KEK Theory Center and the International Visitor
Program for hospitality and the opportunity to present this work.  PM
and GS were supported in part of NSF CAREER Award No. Phy-0348093, DOE
grant DE-FG-02-95ER40896, a Cottrell Scholar Award from Research
Corporation, a Vilas Associate Award from the University of Wisconsin,
and a John Simon Guggenheim Memorial Foundation Fellowship.  YS was
supported by Nishina Memorial Foundation.  GS also would like to
acknowledge support from the Ambrose Monell Foundation during his stay
at the Institute for Advanced Study.

\newpage

\appendix

%%%%%
\section{{\label{sec:notations}}Conventions}
%%%%%

We work in the type IIB supergravity limit where the bosonic part of
the Einstein frame action is \cite{Polchinski:2000uf}
\begin{align}
  \label{eq:iibaction}
  S =& \frac{1}{2 \kappa^2} \int d^{10} x \sqrt{-g} \left[R
    -\frac{1}{2} \partial_M \Phi \partial^M \Phi -
    \frac{1}{2} e^{2 \Phi} \partial_M C \partial^M C  \right.\notag\\
  & \phantom{\frac{1}{2 \kappa^2} \int d^{10} x
    \sqrt{-g}\biggl[}\left. - \frac{g_s}{2\times 3!} e^{-\Phi} H_3^2 -
    \frac{g_s}{2\times 3!} e^{\Phi} \tilde{F}_3^2 -
    \frac{g_s^2}{4\times 5!} {\tilde{F}}_5^2\right]\notag\\
  & - \frac{g_s^2}{4 \kappa^2} \int C_4 \wedge H_3 \wedge F_3,
\end{align}
where we use
\begin{equation}
  \tilde{F}_3 = d C_2 - C H_3, \quad
  \tilde{F}_5 = d C_4 + B_2 \wedge F_3 = (1+*_{10}) {\cal F}_5, \quad
  2\kappa^2 = (2\pi)^7 \alpha'^4 g_s^2,
\end{equation}
and the self-duality of $\tilde{F}_{5}$ is imposed at the level of the
equations of motion.  The string frame metric is related to the
Einstein frame metric by $g^{E}_{MN} = g_s^{\frac{1}{2}}
e^{-{\frac{\Phi}{2}}} g^s_{MN}$.

The equations of motion resulting from~\eqref{eq:iibaction} are
\begin{subequations}
  \label{eom}
  \begin{align}
    &R_{MN} -\frac{1}{2} g_{MN} R - \frac{1}{2} \partial_M
    \Phi \partial_N\Phi - \frac{1}{2} e^{2\Phi} \partial_M
    C \partial_N C  \notag\\
    &\quad -\frac{g_s}{2\times 2!}e^{-\Phi}
    H_{MR_1R_2}H_{N}^{\phantom{M}R_1R_2} - \frac{g_s}{2\times 2!}
    e^{\Phi}\tilde{F}_{MR_1R_2}
    \tilde{F}_{N}^{\phantom{N}R_1R_2}  \notag \\
    &\quad -\frac{g_s^2}{4 \times 4!}\tilde{F}_{MR_1R_2R_3R_4}
    \tilde{F}_{N}^{\phantom{N}R_1R_2R_3R_4} \notag\\
    &\quad +\frac{1}{2}g_{MN}\left[\frac{1}{2} \left(\partial
        \Phi\right)^2 + \frac{1}{2} e^{2\Phi}(\partial C)^2 +
      \frac{g_s}{2\times 3!} e^{-\Phi} H_3^2 + \frac{g_s}{2\times 3!}
      e^{\Phi} \tilde{F}_3^2 +
      \frac{g_s^2}{4\times 5!} {\tilde{F}}_5^2\right]=0,\label{eq:einstein}\\
    &\nabla^2 \Phi - e^{2\Phi} \left(\partial_M C\right)^2 + \frac{g_s
      e^{-\Phi}}{2 \times 3!}  \left[H_3^2 - e^{2\Phi} \tilde{F}_3^2
    \right]=0,
    \label{eq:dilatoneom}\\
    &d *\left(e^{-\Phi} H_3 - C e^{\Phi} \tilde{F}_3\right) +
    g_s F_5 \wedge F_3 = 0,\label{eq:H3 original eom}\\
    &d * \left(e^{\Phi} \tilde{F}_3\right) - g_s F_5 \wedge H_3 = 0,\label{eq:F3 original eom}\\
    &d *{\tilde{F}_5} - H_3\wedge F_3 = 0 \label{eq:F5 eom}.
  \end{align}
\end{subequations}
Note that imposing the self-duality of $\tilde{F}_{5}$ implies
$\tilde{F}_{5}^{2}=0$.  With the ansatz~(\ref{geometry
  ansatz}),~(\ref{gauge ansatz}), and taking
$\ell=f\left(1-F\right)+kF$, the Bianchi identity for $\tilde{F}_5$ is
automatically satisfied.
With this ansatz, the equations for $H_{3}$
(\ref{eq:H3 original eom}) can be written
\begin{subequations}
  \begin{align}
    &\frac{d}{d\tau}\left(e^{-\Phi}h^{-1}
      \sqrt{\frac{b}{p}}\frac{q}{s}f'\right)+
    \frac{e^{-\Phi}}{2h}\sqrt{\frac{p}{b}} (k-f) - \frac{g_s M^2
      \alpha'^2}{4 h^2}\sqrt{\frac{p}{b}} \frac{\ell (1-F)}{q s}=0,
    \label{H3eom 1}\\
    &\frac{d}{d\tau}\left(e^{-\Phi}h^{-1}\sqrt{\frac{b}{p}}\frac{s}{q}
      k'\right) - \frac{e^{-\Phi}}{2h}\sqrt{\frac{p}{b}} (k-f) -
    \frac{g_s M^2 \alpha'^2}{4 h^2}\sqrt{\frac{p}{b}} \frac{\ell F}{q
      s}=0,
    \label{H3eom 2}
  \end{align}
\end{subequations}
while the equation for $F_3$ (\ref{eq:F3 original eom}) is
\begin{align}
  \frac{d}{d\tau}\left(e^{\Phi} h^{-1} \sqrt{\frac{b}{p}} F' \right) +
  \frac{e^{\Phi}}{2h} \sqrt{\frac{p}{b}} \left[(1-F) \frac{s}{q} -F
    \frac{q}{s}\right] - \frac{g_s^3 M^2 \alpha'^2}{8 h^2}
  \sqrt{\frac{p}{b}} \frac{\ell (k-f)}{q s} = 0.
  \label{F3eom}
\end{align}

The bosonic and fermionic actions together are invariant under the
supersymmetric transformations for the gravitino $\Psi_{M}$ and
dilatino $\lambda$,
\begin{subequations}
  \label{susy variations}
  \begin{align}
    \delta \Psi_M =& D_M \epsilon + i \frac{\sqrt{g_s} e^{\Phi /
        2}}{96} \left( \Gamma_{M}^{\phantom{M}R_1R_2R_3} G_{R_1R_2R_3}
      -
      9\Gamma^{R_1R_2} G_{M R_1R_2}\right)\epsilon^*  \notag \\
    &+i\frac{g_s}{192}\Gamma^{R_1R_2R_3R_4} F_{MR_1R_2R_3R_4}
    \epsilon,
    \label{eq:gravitinovar}\\
    \delta \lambda =& \frac{i}{2} \Gamma^R \left(i e^\Phi \partial_R C
      +
      \partial_R \Phi\right)\epsilon^* - \frac{e^{\Phi}}{2}
    \Gamma^R \partial_R C \epsilon + \frac{\sqrt{g_s} e^{\Phi /
        2}}{24} \Gamma^{R_1R_2R_3} G_{R_1R_2R_3}\epsilon,
    \label{eq:dilatinovar}
  \end{align}
\end{subequations}
together with accompanying bosonic transformations. Here,
\begin{equation}
  G_3 \equiv F_3 - \tau_{ad} H_3,\quad
  F_{3}=dC_{2},\quad
  \tau_{ad}\equiv C_{0}+ie^{-\Phi}.
\end{equation}

We can use these transformations to check if supersymmetry is
respected by the solution. Using the ansatz~\eqref{geometry
  ansatz},~\eqref{gauge ansatz}, the supersymmetry conditions
$\delta\Psi_{M}=\delta\lambda=0$ imply
\begin{align}
  \label{ISD & IASD condition}
  1-F - g_s e^{-\Phi}\sqrt{\frac{b}{p}}\frac{q}{s}f'=&0,\quad F - g_s
  e^{-\Phi} \sqrt{\frac{b}{p}}\frac{s}{q} k'=0,\quad F' -
  \frac{g_s}{2} e^{-\Phi} \sqrt{\frac{p}{b}} (k-f)=0,
\end{align}
which impose that the flux $G_{3}$ is imaginary-self-dual (ISD).  One
can further show that supersymmetry requires that the flux be a
primitive $\left(2,1\right)$-form.  The variation for the
gravitino~\eqref{eq:gravitinovar} requires the warp factor to be
related to $F_5$,
\begin{equation}
  \label{eq:warpfactorsusy}
  h' = - \frac{(g_s M \alpha')^2}{4}\frac{\ell}{q s}
  \sqrt{\frac{p}{b}}.
\end{equation}
This condition implies that the BPS condition equates the tension and
charge of $3$-branes added to the geometry.

The conifold and its related geometries make use of the angular
$1$-forms
\begin{align}
  &e_1= - \sin \theta_1 d\phi_1,\quad e_2 = d\theta_1, \quad
  e_3 = \cos \psi  \sin \theta_2 d\phi_2 - \sin \psi d\theta_2,\notag\\
  &e_4 = \sin \psi \sin \theta_2 d\phi_2 + \cos \psi d\theta_2,\quad
  e_5 = d\psi + \cos \theta_1 d\phi_1 + \cos \theta_2 d\phi_2.
\end{align}
In terms of these it is also useful to define~\cite{Minasian:1999tt}
\begin{align}
  &g_1 = \frac{e_1- e_3}{\sqrt{2}}, \quad g_2 = \frac{e_2 -
    e_4}{\sqrt{2}},\quad g_3 = \frac{e_1 + e_3}{\sqrt{2}}, \quad g_4 =
  \frac{e_2 + e_4}{\sqrt{2}},\quad g_5 = e_5,
\end{align}
which satisfy
\begin{align}
  &d(g_1 \wedge g_3 + g_2\wedge g_4)=g_5\wedge (g_1\wedge g_2 - g_3\wedge g_4),\notag\\
  &d(g_1 \wedge g_2 - g_3\wedge g_4)=-g_5\wedge (g_1\wedge g_3 + g_2\wedge g_4),\notag\\
  &d(g_1\wedge g_2 + g_3 \wedge g_4)=0,\notag\\
  &d g_5 \wedge g_1 \wedge g_2 = d g_5 \wedge g_3 \wedge g_4 = 0,\notag\\
  &d(g_5 \wedge g_1\wedge g_2)= d(g_5\wedge g_3\wedge g_4) = 0.
\end{align}

%%%%
\section{Complex Coordinates\label{sec:complex-coordinates}}
%%%%

The angular coordinates and radial coordinate of the deformed conifold
are related to the complex coordinates $z_{i}$
by~\cite{Candelas:1989js}
\begin{subequations}
  \begin{align}
    W =& L_1 \cdot W_0 \cdot L_2^\dagger\equiv
    \begin{pmatrix}
      z_3 + i z_4& z_1 - i z_2\\
      z_1+i z_2 & -z_3 + i z_4
    \end{pmatrix},\\
    L_j =& \begin{pmatrix} \cos \frac{\theta_j}{2} \, e^{i(\psi_j +
        \phi_j)/2}&
      -\sin\frac{\theta_j}{2} \, e^{-i(\psi_j - \phi_j)/2}\\
      \sin \frac{\theta_j}{2}\,e^{i(\psi_j - \phi_j)/2}& \cos
      \frac{\theta_j}{2} \, e^{-i(\psi_j + \phi_j)/2}
    \end{pmatrix},\quad W_0 = \begin{pmatrix}
      0&	\varepsilon e^{\tau/2}\\
      \varepsilon e^{-\tau/2} & 0
    \end{pmatrix},
  \end{align}
\end{subequations}
and the $z_{i}$ satisfy
\begin{equation}
  \sum_{i=1}^{4}z_{i}^{2}=\varepsilon^{2}.
\end{equation}
The angles $\psi_{i}$ always appear in the combination
$\psi=\psi_{1}+\psi_{2}$.  For $\varepsilon\neq 0$, %the radial coordinate
$\tau$ is defined by
\begin{equation}
  R^2 = \sum_{i=1}^4 z_i \bar{z}_i = \frac{1}{2}
  \Tr\left(W\cdot W^\dagger\right) = \varepsilon^2 \cosh \tau.
\end{equation}

The deformed conifold metric can be written as \cite{Candelas:1989js}
\begin{subequations}
\begin{align}
  ds_6^2 =& \partial_i \partial_{\bar{\jmath}} {\cal F} \, dz_i d
  \bar{z}_j\notag \\
  =&\frac{1}{4}{\cal F}''(R^2) \left|\Tr \left(W^\dagger
      dW\right)\right|^2 +
  \frac{1}{2} {\cal F}'(R^2) \Tr\left(dW^\dagger dW\right)\notag \\
  =& -i J_{i\bar{\jmath}}dz_i d\bar{z}_{\bar{\jmath}},
\end{align}
where
\begin{align}
  J =& j_{dc}(\tau) (g_2\wedge g_3 + g_4 \wedge g_1) +
  d j_{dc}(\tau) \wedge g_5,\notag \\
  {\cal F}'(R^2) =& \varepsilon^{-\frac{2}{3}} K(\tau),\notag \\
  j_{dc}(\tau) =& \frac{\varepsilon^2}{2} \sinh \tau {\cal F}'(R^2),
  \label{complex form of metric and almost complex structure}
\end{align}
\end{subequations}
and where $'$ indicates a derivative with respect to $R^{2}$ and $J$ is
the almost complex structure.

%\newpage

It is convenient to write $G_{3}$ in terms of these complex
coordinates.  Following~\cite{Herzog:2001xk}, we consider the $SO(4)$
invariant $1$-forms and $2$-forms
\begin{align}
  &\xi_1=\bar{z}_i dz_i,\quad \xi_2=z_i d\bar{z}_i,\notag\\
  &\eta_1 = \epsilon_{ijkl} z_i \bar{z}_j dz_k \wedge d\bar{z}_l,\quad
  \eta_2 = \epsilon_{ijkl} z_i \bar{z}_j dz_k \wedge dz_l,\quad
  \eta_3 = \epsilon_{ijkl} z_i \bar{z}_j d\bar{z}_k \wedge d\bar{z}_l,\notag\\
  &\eta_4 = (z_i d\bar{z}_j)\wedge (\bar{z}_j dz_j),\quad \eta_5 =
  dz_i \wedge d\bar{z}_i.
\end{align}
In terms of these,
\begin{subequations}
  \begin{align}
    d\tau =& \frac{1}{\ve^2\sinh\tau} \bigl(z_{i}d\bar{z}_{i} +
    \bar{z}_{i}d z_{i}\bigr),\qquad g_{5} =
    \frac{i}{\ve^2\sinh\tau}\bigl(z_{i}d\bar{z}_{i} -
    \bar{z}_{i}d z_{i}\bigr),\\
    g_{1}\wedge g_{2}=& \frac{i\bigl(1+\cosh\tau)}
    {2\ve^{4}\sinh^{3}\tau}\epsilon_{ijkl}\bigl( 2z_{i}\bar{z}_{j}d
    z_{k}\wedge d\bar{z}_{l}- z_{i}\bar{z}_{j}d z_{k}\wedge d z_{l}-
    z_{i}\bar{z}_{j}d\bar{z}_{k}\wedge d\bar{z}_{l} \bigr),\\
    g_{3}\wedge g_{4}=&\frac{i\tanh\frac{\tau}{2}}
    {2\ve^{4}\sinh^{2}\tau}\epsilon_{ijkl}\bigl( 2z_{i}\bar{z}_{j}d
    z_{k}\wedge d\bar{z}_{l}+ z_{i}\bar{z}_{j}d z_{k}\wedge d z_{l} +
    z_{i}\bar{z}_{j}d\bar{z}_{k}\wedge d\bar{z}_{l} \bigr),\\
    g_{1}\wedge g_{3} + g_{2} \wedge g_{4} =&
    \frac{1}{\ve^{4}\sinh^{2}\tau}\epsilon_{ijkl}\bigl(
    -z_{i}\bar{z}_{j}d z_{k}\wedge d z_{l} +
    z_{i} \bar{z}_{j} d\bar{z}_{k} \wedge d\bar{z}_{l}\bigr),\\
    g_{2}\wedge g_{3} + g_{4}\wedge g_{1} =&
    -\frac{2i\cosh\tau}{\ve^{4}\sinh^3\tau} \bigl(\bar{z}_{j}d
    z_{j}\bigr)\wedge \bigl(z_{i}d\bar{z}_{i}\bigr) +
    \frac{2i}{\ve^2\sinh\tau}d z_{i} \wedge d\bar{z}_i.
  \end{align}
\end{subequations}

The other remaining $1$-forms cannot be as easily written in terms of
the complex coordinates.  However, we find
\begin{subequations}
  \begin{align}
    g_1^2 + g_2^2 =& -\frac{1}{2 \varepsilon^4 \sinh^2 (\tau/2)
      \sinh^2\tau} \left[(\bar{z}\cdot dz)^2 + (z\cdot d\bar{z})^2 +
      2\cosh\tau (\bar{z}\cdot dz)(z\cdot d\bar{z}) \right.\notag\\
    &\hspace{11em} \left.+ \varepsilon^2 \sinh^2 \tau (dz\cdot dz+
      d\bar{z} \cdot d\bar{z} - 2 dz\cdot d\bar{z}) \right],\notag\\
    g_3^2 + g_4^2 =& \frac{1}{2 \varepsilon^4 \cosh^2 (\tau/2)
      \sinh^2\tau} \left[ (\bar{z}\cdot dz)^2 + (z\cdot d\bar{z})^2 -
      2\cosh\tau (\bar{z}\cdot dz)(z\cdot d\bar{z}) \right.\notag\\
    &\hspace{11em} \left.+ \varepsilon^2 \sinh^2 \tau (dz\cdot dz+
      d\bar{z} \cdot d\bar{z} + 2 dz\cdot d\bar{z}) \right].
  \end{align}
\end{subequations}

In terms of these complex coordinates
\begin{subequations}
  \label{(0,3) and (3,0)-forms}
  \begin{align}
    G_{3}^{\left(3,0\right)}= \frac{M\alpha'}{2\ve^{6}}
    \biggl[&\left\{\left(1-F\right)
      \frac{\tanh\frac{\tau}{2}}{2\sinh^{3}\tau}
      -F\frac{1+\cosh\tau}{2\sinh^{4}\tau}
      -\frac{F'}{\sinh^{3}\tau}\right\}\biggr. \notag \\
    &\biggl.+g_{s}e^{-\Phi}\left\{
      -f'\frac{1+\cosh\tau}{2\sinh^{4}\tau}
      +k'\frac{\tanh\frac{\tau}{2}}{2\sinh^{3}\tau}
      +\frac{k-f}{2\sinh^{3}\tau}\right\}\biggr]
    \xi_{1}\wedge \eta_{2}, \\
    \label{eq:03component}
    G_{3}^{\left(0,3\right)}= \frac{M\alpha'}{2\ve^{6}}
    \biggl[&-\left\{\left(1-F\right)
      \frac{\tanh\frac{\tau}{2}}{2\sinh^{3}\tau}
      -F\frac{1+\cosh\tau}{2\sinh^{4}\tau}
      -\frac{F'}{\sinh^{3}\tau}\right\}\biggr. \notag \\
    &\biggl.+g_{s}e^{-\Phi}\left\{
      -f'\frac{1+\cosh\tau}{2\sinh^{4}\tau}
      +k'\frac{\tanh\frac{\tau}{2}}{2\sinh^{3}\tau}
      +\frac{k-f}{2\sinh^{3}\tau}\right\}\biggr]
    \xi_{2}\wedge\eta_{3}.
  \end{align}
\end{subequations}
For the KS solution~\eqref{KS solution}, these components vanish since
each of the terms in braces vanishes independently.  The remaining
components of $G_{3}$ are
\begin{subequations}
  \begin{align}
    G_{3}^{\left(2,1\right)}=&\frac{M\alpha'}{2\ve^{6}} \biggl\{
    2\bigl(a_{1}^{+}+a_{2}^{+}\bigr)\xi_{1}\wedge\eta_{1} +
    \bigl(a_{1}^{-}-a_{2}^{-}-a_{3}^{+}\bigr)
    \xi_{2}\wedge\eta_{2}\biggr\}\\
    G_{3}^{\left(1,2\right)}=&\frac{M\alpha'}{2\ve^{6}} \biggl\{
    2\bigl(a_{1}^{-}+a_{2}^{-}\bigr)\xi_{2}\wedge\eta_{1} +
    \bigl(a_{1}^{+}-a_{2}^{+}-a_{3}^{-}\bigr)
    \xi_{1}\wedge\eta_{3}\biggr\}
  \end{align}
\end{subequations}
where we have defined
\begin{subequations}
  \begin{align}
    a_{1}^{\pm}\left(\tau\right) =&
    \frac{\tanh\frac{\tau}{2}}{2\sinh^{3}\tau}
    \left(\pm\left(1-F\right)+g_{s}e^{-\Phi}k'\right), \\
    a_{2}^{\pm}\left(\tau\right) =&\frac{1+\cosh\tau}{2\sinh^{4}\tau}
    \left(\pm F + g_{s}e^{-\Phi}f'\right), \\
    a_{3}^{\pm}\left(\tau\right) =&\frac{1}{\sinh^{3}\tau} \left(\pm
      F'+g_{s}e^{-\Phi}\frac{k-f}{2}\right).
  \end{align}
\end{subequations}
For the KS solution, the only non-vanishing term is the
$\left(2,1\right)$-form.  The $3$-form flux for the KS solution can
also be shown to satisfy the primitivity condition $G_{3}\wedge J=0$.

In calculating the gravitino mass, we make use of the holomorphic
$\left(3,0\right)$-form of the deformed conifold.
Explicitly~\cite{Dymarsky:Thesis,Maldacena:2009mw,Dymarsky:2009fj},
\begin{align}
  \Omega =& \frac{\varepsilon^2}{16 \sqrt{3}} \left[-\sinh \tau
    (g_1\wedge g_3 + g_2 \wedge g_4 ) +
    i \cosh \tau (g_1\wedge g_2 - g_3\wedge g_4) \right. \notag\\
  & \phantom{\frac{\varepsilon^2}{16 \sqrt{3}}\bigl[} \left. - i
    (g_1\wedge g_2+ g_3\wedge g_4)\right]
  \wedge (d\tau + i g_5) \notag\\
  =& \frac{1}{4 \sqrt{3} \,\varepsilon^4 \sinh^2 \tau}
  (\epsilon_{ijkl} z_i \bar{z}_j d{z}_k \wedge d{z}_l) \wedge
  (\bar{z}_m dz_m).
  \label{holomorphic 3-form}
\end{align}
$\Omega$ is normalized so that $\Omega \wedge \bar{\Omega} /
||\Omega||^2 = i {\rm vol}_6$ with $ ||\Omega||^2 \equiv \Omega_{ijk}
\bar{\Omega}^{ijk}/3! = 1$ where the indices are contracted with the
unwarped metric.  The holomorphic $3$-form and the almost complex
structure~(\ref{complex form of metric and almost complex structure})
satisfy the algebraic constraints $\Omega \wedge \bar{\Omega} = -
\frac{i}{48} J^3$, $J\wedge \Omega =0$ and the sourceless calibration
conditions, $d\Omega = d J = d (J\wedge J) =0$.

Some of these expressions simplify if we adopt an alternative basis of
holomorphic $1$-forms,
\begin{align}
  dZ_1 \equiv d\tau + i g_5,\quad dZ_2 \equiv g_1 - i \coth
  \frac{\tau}{2}\, g_4,\quad dZ_3 \equiv g_3 - i \tanh
  \frac{\tau}{2}\, g_2.
\end{align}
In these coordinates, $G_{3}$ is
\begin{subequations}
  \begin{align}
    G_3 =& -\frac{M \alpha'}{16 \sinh^2 \tau} \left[4(\sinh \tau -
      \tau \cosh\tau)d\bar{Z}_1 \wedge dZ_2\wedge dZ_3
    \right. \notag\\
    &\left. \hspace{6em} + (\sinh 2\tau -2\tau) \left(d\bar{Z}_2\wedge
        dZ_1\wedge dZ_3 +
        d\bar{Z}_3 \wedge dZ_1 \wedge dZ_2 \right)\right],\\
    \intertext{while the holomorphic 3-form and metric for the
      deformed conifold are} \Omega =& - \frac{\varepsilon^2}{16
      \sqrt{3}}\sinh\tau\,
    dZ_1\wedge dZ_2\wedge dZ_3,\\
    ds_6^2 =& \frac{\varepsilon^{4/3}}{6 K^2} dZ_1 d\bar{Z}_1 +
    \frac{\varepsilon^{4/3} K}{2} \sinh^2\frac{\tau}{2}\, dZ_2
    d\bar{Z}_2 + \frac{\varepsilon^{4/3} K}{2}\cosh^2 \frac{\tau}{2}\,
    dZ_3 d\bar{Z}_3
  \end{align}
\end{subequations}
where $K$ is defined in (\ref{KS solution}).

\bibliographystyle{JHEP}
\bibliography{MSS-sol}

\end{document}